\documentclass[useAMS,usenatbib]{mn2e}

\usepackage{graphicx, subfigure, fixltx2e, placeins}
\usepackage{float}
\usepackage{deluxetable}
\usepackage{amssymb}
\usepackage{longtable}
\usepackage{textcomp}
\usepackage{url}
\usepackage{footnote}

\newcommand{\Msun}{\ensuremath{M_{\odot}}}
\newcommand{\Lsun}{\ensuremath{L_{\odot}}}

\title[Submm/mm Galaxy Counterpart Identification]{Submm/mm Galaxy Counterpart Identification Using a Characteristic Density Distribution}
\author[Alberts et al.]{\parbox{\textwidth}{
Stacey Alberts$^1$, 
Grant W. Wilson$^1$, 
Yu Lu$^2$,
Seth Johnson$^1$, 
Min S. Yun$^1$, 
Kimberly S. Scott$^3$,
Alexandra Pope$^1$,
Itziar Aretxaga$^4$, 
Hajime Ezawa$^5$, 
David H. Hughes$^4$,
Ryohei Kawabe$^6$,  
Sungeun Kim$^7$,
Kotaro Kohno$^{8,9}$,
Tai Oshima$^{10}$} \vspace{0.2cm}\\
  $^1$Department of Astronomy, LGRT-B 619E, University of Massachusetts, Amherst, MA 01003, USA \\
  $^2$Kavli Institute for Particle Astrophysics and Cosmology, Stanford, CA 94309, USA \\
  $^3$North American ALMA Science Center, National Radio Astronomy Observatory, Charlottesville, VA 22903, USA \\
  $^4$Instituto Nacional de Astrof\'{i}sica, \'{O}ptica y Electr\'{o}nica, Calle Luis Enrique Erro 1, Sta. Ma. Tonantzintla, Puebla, Mexico \\
  $^5$ALMA Project Office, National Astronomical Observatory of Japan, 2-21-1 Osawa, Mitaka, Tokyo 181-8588, Japan \\
  $^6$Joint ALMA Observatory, 
Alonso de Cordova 3107 OFC 129, Vitacura, Santiago 763 0355, Chile \\
  $^7$Astronomy and Space Science Department, Sejong University, Seoul, South Korea \\
  $^8$Institute of Astronomy, University of Tokyo, 2-21-1 Osawa, Mitaka, Tokyo 181-0015, Japan \\
  $^9$Research Center for the Early Universe, University of Tokyo, 7-3-1 Hongo, Bunkyo, Tokyo 113-0033, Japan \\
  $^{10}$Nobeyama Radio Observatory, National Astronomical Observatory of Japan, Minamimaki, Minamisaku, Nagano 384-1305, Japan 
  }
\date{Updated \today}
  
\pagerange{\pageref{firstpage}--\pageref{lastpage}} \pubyear{2013}
\begin{document}

\label{firstpage}
\maketitle

\begin{abstract}
We present a new submm/mm galaxy counterpart identification technique which builds on the use of $Spitzer$ IRAC colors as discriminators between likely counterparts and the general IRAC galaxy population.  Using 102 radio- and SMA-confirmed counterparts to AzTEC sources across three fields (GOODS-N, GOODS-S, and COSMOS), we develop a non-parametric IRAC color-color characteristic density distribution (CDD), which, when combined with positional uncertainty information via likelihood ratios, allows us to rank all potential IRAC counterparts around SMGs and calculate the significance of each ranking via the reliability factor.  We report all robust and tentative radio counterparts to SMGs, the first such list available for AzTEC/COSMOS, as well as the highest ranked IRAC counterparts for all AzTEC SMGs in these fields as determined by our technique.  We demonstrate that the technique is free of radio bias and thus applicable regardless of radio detections.  For observations made with a moderate beamsize ($\sim$18$^{\prime\prime}$), this technique identifies $\sim$85 per cent of SMG counterparts.  For much larger beamsizes ($\gtrsim$30$^{\prime\prime}$), we report identification rates of 33-49 per cent.  Using simulations, we demonstrate that this technique is an improvement over using positional information alone for observations with facilities such as AzTEC on the LMT and SCUBA-2 on JCMT. 
\end{abstract}


\section{Introduction}
\label{sec:intro}

The discovery of a large population of bright sources at high redshift through the opening of the submillimetre (submm) and millimeter (mm) wavelength windows continues to have a profound impact on our understanding of galaxy evolution in the early Universe.  These submm/mm-selected galaxies (hereafter SMGs) are characterized by large far-infrared luminosities ($\gtrsim$10$^{12}$ $\Lsun$), tremendous star formation rates ($\gtrsim$~500~$\Msun$ yr$^{-1}$), and a number density that is high compared to local ultra-luminous infrared galaxies (ULIRGs), indicating strong evolution within the population \citep[e.g][]{sco02}.  SMGs are thought to represent young, massive systems that exist in an epoch of rapid mass build-up and may be the predecessors of modern-day giant elliptical galaxies \citep[see review by ][]{bla02}.  As a population, SMGs contribute a significant portion of the cosmic energy density at $z$=2-3 through some combination of dusty starbursts and active galactic nuclei \citep[AGN;][]{hug98, ale05, cha05, kov06, pop06, cop08, mur11}.  Observations at submm/mm wavelengths benefit from a strong negative k-correction and thus present a unique opportunity for an unbiased view of star formation out to extreme redshifts.  Understanding SMGs is vital to any complete view of galaxy evolution.  

Currently, thousands of SMGs have been detected in both ground-based surveys at 450-2000~$\mu$m  \citep[e.g.][]{sma97, bar98, hug98, cow02, sco02, bor03, gre04, lau05, cop06, sco08, per08, aus10} and in balloon- and space-based surveys at 250-500~$\mu$m \citep{pas08,dev09,eal10,oli10}.  A complete understanding of the physical processes within these systems, however, requires that these observations be matched to multi-wavelength data, from radio to X-ray.
Such multi-wavelength counterpart identification of SMGs is difficult due to the poor angular resolution ($>$10$^{\prime\prime}$) of single-dish submm telescopes and the intrinsic faintness of their optical counterparts due to dust obscuration.  A great deal of work has been invested into counterpart identification techniques, primarily using radio continuum data from interferometric observations \citep[e.g.][]{ivi02, cha05, pop06, ivi07, cha09, yun12}.  Radio observations are particularly useful for counterpart identification due to 1) the low number density of radio sources, which minimizes the probability of a chance association, and 2) the tight correlation between radio and submm/mm emission \citep[e.g.][]{con92, yun01}, both star formation tracers.  Direct submm/mm interferometric imaging of SMGs with high angular resolution \citep[e.g.][]{you07, you09, smo12, bar12} supports this radio-submm association.

%The use of radio images for counterpart identification is limited, however,  due to some fraction of SMGs which lack a detectable radio counterpart, even in the deepest radio surveys.  
Unfortunately, the use of radio detections for counterpart identification has drawbacks.  Due to the radio-submm correlation, radio counterparts will preferentially be found for the brightest SMGs, which may not be representative of the full SMG population.  Interferometric observations have shown that a small percentage ($\sim$7-15$\%$) of single radio sources near SMGs are not actually associated with the submm emission.  In cases where multiple radio detections are found within the submm beam, interferometry reveals that in $\sim$80$\%$ of these instances, the submm emission is only associated with one of the radio sources \citep[see Section~\ref{sec:posun};][]{you07, you09, smo12, bar12}. Finally, radio observations suffer from a strong positive k-correction, which results in a bias against high redshift galaxies, while submm emission remains equally detectable out to $z$$\sim$8 \citep{bla02}. Current radio observations that overlap with submm surveys have placed the fraction of SMGs that have a detectable radio counterpart at $\sim$40-100 per cent \citep{ivi02, ivi07, wei09, cha09, lin11, yun12, smo12, bar12}, depending on the depth of the radio observations.  For example, given the shallowness of the available radio data in the COSMOS field (see Section~\ref{sec:data}), we can predict that we will detect $\sim$40 per cent of SMGs in the radio based on submm counterpart studies of GOODS-N SMGs, which utilized deep radio data to build a cumulative distribution of radio counterparts as a function of radio depth \citep{pop07}. And, in fact, we find that 40 per cent of SMGs detected with single-dish observations in COSMOS are associated with radio detections \citep[][this work]{are11}.  This is further confirmed by interferometric observations of AzTEC SMGs in COSMOS.  \citet{you07, you09} imaged a sample of 15 SMGs with the Submillimetre Array (SMA) and found that only 6/15 (40$\%$) have a robust radio counterpart.  Of the nine with no radio counterpart, seven are also undetected with MIPS 24~$\mu$m.  On the other hand, 13/15 of these SMGs are detected with $Spitzer$ Infrared Array Camera \citep[IRAC;][]{faz04}.  Of the two not detected, one is confused with a foreground source.  Additionally, several studies have shown that SMGs are  routinely detected in the IRAC bands at 3.6 and 4.5~$\mu$m at the $\ge$ 1~$\mu$Jy level \citep{ion06, wan07, wan11, hat10, tam10, ika11}.  The high source density \citep[$\sim$60 arcmin$^{-2}$ at the 1.4~$\mu$Jy level at 3.6~$\mu$m;][]{faz04b} of IRAC galaxies makes it impossible to use IRAC detections directly for unique identifications \citep{pop06}, however, the selection of IRAC counterparts is expected to suffer less bias against high redshift sources and may be more representative of faint SMGs (see Section~\ref{sec:aztecflux}).

More recently, two studies have examined follow-up interferometry of SMGs detected with LABOCA \citep[870~$\mu$m; ][]{sir09} and SCUBA/SCUBA-2 (850~$\mu$m).  \citet{smo12} analysed a set of 34 LABOCA sources in COSMOS with interferometry from various sources, including 26 new detections from the IRAM Plateau de Bure Interferometer (PdBI).  Using a radio catalog extended to include lower signal-to-noise (S/N) sources, they found that 18/34 ($\sim$50$\%$) SMGs were detected at 20cm and the same fraction, though a different subset, were detected with IRAC.  This study also determines redshifts to their sources and concludes that the redshift distribution of SMGs may peak higher than previously thought, including a large tail of sources at very high redshift ($z$$>$3) that will be difficult to detect in even the deepest radio surveys due to the k-correction. It should be noted, however, that this study did not require the LABOCA SMGs to be detected at high significance.  If we limit the sample to high significance interferometric detections ($>$4.5$\sigma$), then 10/16 (62$\%$) SMGs are detected in IRAC, while, if we look only at LABOCA sources which are also detected with AzTEC, 8/10 (80$\%$) have IRAC counterparts.  Additionally, \citet{bar12} looked at a small, but homogenous sample of 12 SMGs in GOODS-N and found 16 independent, high significance detections with the SMA.  Of these 16, all are detected in an ultra-deep radio catalog and 15 are detected in the IRAC bands.

In this study, we develop a new technique for counterpart identification which takes advantage of the high sensitivity of $Spitzer$ IRAC and does not rely on radio or MIPS 24$\mu$m observations.  Our starting point is the IRAC color-color space diagram first developed by \citet{lac04} for identifying AGN.  IRAC colors, and this diagram in particular, have been shown to also be useful for SMG counterpart identification, with $\lesssim$20 per cent of SMGs overlapping with AGN in IRAC color-color space \citep{pop06, big11, yun08, yun12}.  We take these studies a step farther and, using a large sample of SMGs with secure radio or SMA counterparts, develop an IRAC color-color density distribution.  This multi-dimensional color prior is combined with positional uncertainty information via the likelihood ratio \citep[LR; ][]{sut92}, providing a method for ranking potential IRAC counterparts around SMGs.  Previous studies utilizing likelihood ratios for counterpart identification can be found in the literature across all wavelength regimes; for example, \citet{cha11} combined priors based on radio, MIPS, and IRAC fluxes and colors with positional information to identify counterparts to BLAST and LABOCA sources.  \citet{smi11} and \citet{kim12} similarly used individual flux and color priors to identify counterparts to $Herschel$ SPIRE sources.  The methods used in these studies necessarily assume that the various priors used are independent, though some or all may be correlated \citep{cha11}.  In this work, we incorporate correlations between different flux bands and colors by using a characteristic density distribution (CDD) rather than individual priors.

Since our method only requires $Spitzer$ IRAC observations, it can be applied in numerous fields that lack deep radio observations and is not as strongly biased against high redshift sources as techniques which rely on radio data.  
Many of the fields in which deep IRAC data already exists will be prime targets for future submm surveys and we show that the technique developed here is useful not only for existing submm surveys, but also for upcoming surveys with the LMT and SCUBA-2.  Though the particular characteristic density distribution developed in this study is only applicable to source populations with a redshift distribution and spectral energy distribution (SED) similar to sources selected at 850-1200~$\mu$m (see Section~\ref{sec:physics}), we posit that a modified version of this technique can be developed for $Herschel$ SPIRE sources, which suffer from similar challenges due to low resolution.

This paper is organized as follows.  In Section~\ref{sec:data}, we describe the data used in this study.  In Section~\ref{sec:id}, we develop the counterpart identification technique, which we use in Section~\ref{sec:counter} to calculate an LR and its corresponding reliability for all IRAC counterpart candidates across the three fields, including an analysis of the resulting ranking via a blind-test using our radio- and SMA-identified counterpart sample.  In Section~\ref{sec:disc}, we discuss our technique in the context of future submm surveys and address the issue of possible bias introduced through the use of radio counterparts in developing the IRAC color-color characteristic density distribution.  Section~\ref{sec:con} contains our conclusions.

\section{Data}
\label{sec:data}

To develop our new technique for selecting IRAC counterparts to SMGs, we utilize a statistically significant sample of 272 SMGs detected at 1.1 mm with AzTEC  \citep{wil08} from three fields: the Cosmic Evolution Survey (COSMOS), the Great Observatories Origins Deep Survey-North (GOODS-N), and -South (GOODS-S). These three fields have extensive multi-wavelength data ranging from X-ray to radio, much of which was observed at high resolution.  We take advantage of observations from the Very Large Array (VLA) at 1.4 GHz and interferometric observations taken with the SMA to identify secure counterparts for a subset of the AzTEC SMGs; the IRAC counterparts are then used to derive a characteristic density distribution utilizing the four $Spitzer$ IRAC bands (3.6, 4.5, 5.8, and 8.0~$\mu$m), which is described in the next sections.  We describe each data set below.

$AzTEC$: The AzTEC observations used here are described in the following papers:  \citet{per08} for GOODS-N, \citet{sco10} for GOODS-S, and \citet{sco08} and \citet{are11} for COSMOS.  Data reduction was performed using the standard AzTEC pipeline  \citep[see][]{sco08,dow11}.  The reader is referred to \citet{dow11} for the most updated source lists for the GOODS fields.  Observations in GOODS-N were taken on the JCMT with uniform coverage to a depth of $\sim$1.3 mJy beam$^{-1}$ and a full width at half max (FWHM) of 18$^{\prime\prime}$.  GOODS-S and COSMOS were imaged using the Atacama submillimetre Telescope Experiment \citep[ASTE;][]{eza04} to a depth of $\sim$0.6 and $\sim$1.3 mJy beam$^{-1}$, respectively.  These observations were done at a lower angular resolution and we measure a FWHM of 30$^{\prime\prime}$ for GOODS-S and 33$^{\prime\prime}$ for COSMOS by fitting the post-filtered PSFs of each set of observations with a Gaussian.  Our sample contains 36, 47, and 189 SMGs in GOODS-N, GOODS-S, and COSMOS, for a total of 272 sources.

$VLA$: Radio observations at 20 cm (1.4 GHz) were obtained for all fields using the VLA with an angular resolution of $\leq$3.5$^{\prime\prime}$.  Our deepest radio catalog is in GOODS-N, with a sensitivity of 3.9~$\mu$Jy beam$^{-1}$ (1$\sigma$) and 1,230 5$\sigma$ radio detections \citep{mor10}.  Our GOODS-S catalog was made using the ``Search and Destroy" (SAD) procedure in AIPS to identify all sources greater than 3$\sigma$, which were then cross-matched with the IRAC SIMPLE and GOODS catalogs to exclude noise peaks \citep[see][]{yun12}.  The GOODS-S catalog has an rms sensitivity of 8.5~$\mu$Jy beam$^{-1}$ and $\sim$1,500 sources.  For COSMOS, we use the $>$4.5$\sigma$ catalog of \citet{sch10}, which has $\sim$3,000 sources to a 1$\sigma$ sensitivity of 10~$\mu$Jy beam$^{-1}$.  We demonstrate in Section~\ref{sec:multiples} that the sensitivity of the radio catalog has little impact on our results.

$SMA:$ Fifteen AzTEC/JCMT sources in the COSMOS field were the targets of two follow-up observations by the SMA \citep{you07,you09}.  All were detected at 890~$\mu$m with positional accuracies of $\sim$0.2$^{\prime\prime}$.  6/15 sources are detected in the radio, while nine are not.  We include the three radio-undetected sources that were detected in all four IRAC bands in our analysis and in our derivation of the IRAC color-color CDD below.

$Spitzer$ $IRAC:$ GOODS-N, GOODS-S, and COSMOS were imaged with all four IRAC bands with an angular resolution of 2$^{\prime\prime}$ and a positional accuracy of $\sim$0.2$^{\prime\prime}$.  In GOODS-S, the SIMPLE survey \citep{dam11} covers $\sim$1,600 arcmin$^{-2}$ to a 5$\sigma$ depth of 1.0, 1.2, 6.0, and 6.7~$\mu$Jy for bands 3.6, 4.5, 5.8, and 8.0~$\mu$m.  The SIMPLE IRAC catalog comprises all 3.6~$\mu$m detections with S/N $>$ 5.    Deeper observations were done with a 5$\sigma$ sensitivity of 0.1, 0.2, 1.5, and 1.6~$\mu$Jy in GOODS-N \citep{dic00, tre06} and shallower observations were taken with depths of 0.9, 1.7, 11.3, and 14.6~$\mu$Jy in S-COSMOS \citep{san07}, with a catalog containing all sources detected at 3.6~$\mu$m at the $\gtrsim$1~$\mu$Jy level.  Based on studies in the GOODS fields, these sensitivities allow for the detection of galaxies with stellar masses greater than 10$^{10}$ $\Msun$ and a typical age of a few hundred million years up to $z$ $\sim$ 6 \citep{yan06}.  

\section{Deriving the Characteristic Density Distribution}
\label{sec:id}

Our submm galaxy counterpart identification technique relies on two pieces of information:  the positional uncertainty of a given AzTEC detection and the general distribution of IRAC colors for the SMG population.  The latter is obtained from our `training set', the largest sample of radio- and SMA-identified counterparts to SMGs selected at 1.1mm used to date in a counterpart identification study.  We describe the selection of secure radio counterparts and the development of the IRAC color-color CDD of SMGs in this section.

\subsection{Identifying Radio Counterparts to SMGs: the Training Set}
\label{sec:posun}

Due to the typically low S/N ($\lesssim$10) and low angular resolution of submm/mm observations with single-dish telescopes, the positional offsets between the detected emission and the true position of the source are large due to noise fluctuations in the the submm maps.  We can characterize these positional uncertainties though simulations wherein simulated sources of known flux densities are inserted into the map one at a time at random positions.  From this we build a cumulative positional probability distribution $P($$>$$D;S/N)$, which is the probability that a source will be detected at a distance greater than D from its true position as a function of S/N, where the noise is the instrument noise of the observation \citep[see][]{per08, sco10, are11}.  The distributions determined from these simulations are consistent with the analytical solution from \citet{ivi07}, where

\begin{equation}
\label{eqn:pu}
P(>D;S/N) = \int D\textquotesingle\exp\left({\frac{-D\textquotesingle^2}{2\sigma^2}}\right)dD\textquotesingle
\end{equation}

with

\begin{equation}
\label{eqn:pu2}
\sigma \simeq \frac{0.6\mbox{FWHM}}{\mbox{S/N}}
\end{equation}
%The simulated positional uncertainties and their comparison to the \citet{ivi07} formula can be seen in Figure~\ref{fig:pu}.

Using this formula, we can identify all potential radio counterparts around each SMG using a variable search radius based on the S/N of the AzTEC detection.  To be conservative, we select all radio sources within a search radius where $P($$>$$D;S/N)$ goes to one per cent.  For the lowest S/N sources considered here (S/N = 3.5), this corresponds to a search radius of 9.5$^{\prime\prime}$, 16$^{\prime\prime}$, and 17.6$^{\prime\prime}$ for SMGs in GOODS-N, GOODS-S, and COSMOS, respectively.  Any AzTEC source that has a single radio source within the appropriate search radius is included in our training set.  From interferometric follow-ups to AzTEC and LABOCA SMGs, we expect that the false association rate for single radio sources detected within the submm beam to be $\sim$7-15$\%$ \citep{you07, you09, smo12}.  Using a variable search radius based on the positional uncertainties, as in this work, should decrease this rate.  An analysis of the twelve radio-detected SMGs in our training set that have interferometric data available (AzTEC/GN3, AzTEC/GN5, AzTEC/GN14, AzTEC/C3, AzTEC/C12, AzTEC/C13, AzTEC/C14, AzTEC/C18, AzTEC/C22, AzTEC/C38, AzTEC/C98, AzTEC/C145) shows that we have selected the correct radio counterpart in all cases except AzTEC/GN14 \citep[see Table~\ref{app:robust}; ][]{you07, you09, smo12, bar12}.  AzTEC/GN14 has a faint radio counterpart that is below the detection limits of our catalog, in addition to an unrelated radio source within its search radius.

Traditionally, the robustness of a radio counterpart is determined by the P-statistic \citep{dow86}, with a P-stat $<$ 0.05 considered a robust counterpart and 0.05 $<$ P-stat $<$ 0.2 considered a tentative counterpart.  We identify a total of 107 isolated radio counterparts to our SMGs, 18 in GOODS-N, 15 in GOODS-S, and 74 in COSMOS.  Of these, 95 (79$\%$) have P-stat $<$ 0.05, while 22 (21$\%$) have 0.05 $<$ P-stat $<$ 0.2.  We elect to include these tentative counterparts in our training set and have found that they do not significantly influence the resulting IRAC color distribution.  Our radio counterparts are in good agreement with those found in \citet{cha09} for GOODS-N and \citet{yun12} for GOODS-S.  This work presents the formal analysis of the radio counterparts to AzTEC/COSMOS sources described in \citet{are11} (Tables~\ref{app:robust}-\ref{app:multiples}).  

IRAC counterparts are obtained by searching within 2" of the radio positions, which is possible due to the excellent positional accuracy of the VLA observations \citep[$<$1$^{\prime\prime}$;][]{kel08}.  For determining the IRAC color-color CDD, we require that IRAC sources be detected in all four bands.  This leaves 96/107 (90$\%$) radio-detected SMGs with complete IRAC data.  Of the remainder, ten are completely undetected and one is detected in only 3 bands.  To this sample, we add an additional six SMGs with secure IRAC counterparts (detected in all four bands) determined from the SMA imaging that are radio-undetected, have multiple radio counterparts, or are not in the current AzTEC/COSMOS catalog. This gives a total of 102 IRAC counterparts which we use to build an IRAC color space CDD.  The list  of radio and IRAC counterparts to radio-detected AzTEC sources\footnote{A list of AzTEC sources and radio/IRAC counterparts, including photometry, is available at \url{www.astro.umass.edu/aztec/CounterpartID/counterpartid.html}.} can be seen in Table~\ref{app:robust}, including the search radii used and associated P-statistics.  Those AzTEC sources not included in the training set due to lack of an IRAC counterpart are indicated.

We find that 8 (22$\%$), 4 (9$\%$), and 15 (8$\%$) SMGs in GOODS-N, GOODS-S, and COSMOS have multiple radio counterparts$^1$ within their search radii (Table~\ref{app:multiples}).  SMGs with multiple radio counterparts are not included in the training set due to ambiguity in the source of the mm emission.

%***\emph{Move to later sections?}***Finally, the variable search radius is used to collect the 2,018 IRAC galaxies around all AzTEC sources.  Those that are part of the robust counterpart list have their surrounding IRAC galaxies used in a blind-test of the final technique.  All SMGs lacking a radio- or SMA-confirmed counterpart have their surrounding IRAC galaxies ranked by the final technique and can be seen in Table~\ref{app:main}.

%Our secondary use for the positional uncertainties is to directly incorporate the the P($>$D;S/N) of each individual potential IRAC counterpart as a positional likelihood.  This is combined with the IRAC colors of each IRAC galaxy via the LR to determine the relative likelihood that it is the actual counterpart.

\subsection{IRAC Color-Color Space}
\label{sec:irac}

\begin{figure}
\includegraphics[scale=0.55, trim=12mm 0 2mm 2mm, clip]{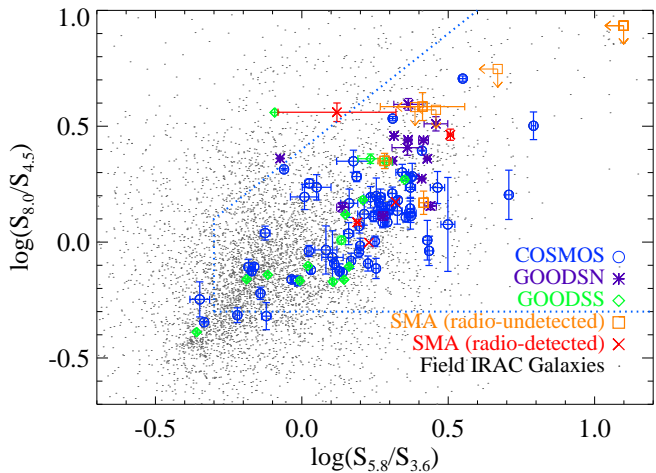}
\caption{The radio- and SMA-identified IRAC counterparts to SMGs that are used as a training set in this study.   In addition to the 102 IRAC counterparts detected in all four IRAC channels, 4 counterparts that are missing one or more IRAC detections are plotted as upper limits. These sources are not part of the training set.  The dotted lines denote the IRAC color cuts suggested for SMG counterpart identification in \citet{yun08}.  The gray dots represent a random subsample of $\sim$10,000 IRAC field galaxies.}
\label{fig:irac}
\end{figure}

Following the lead of \citet{lac04} and \citet{yun08, yun12}, we create an IRAC color-color diagram with the colors $\log\left(\frac{S_{5.8}}{S_{3.6}}\right)$ versus $\log\left(\frac{S_{8.0}}{S_{4.5}}\right)$.
The 102 IRAC counterparts in our training set can be seen in IRAC color-color space in Figure~\ref{fig:irac}, along with a random sample of 10,000 IRAC galaxies from our three IRAC catalogs. The dotted lines represent the color cuts presented in \citet{yun08}, which encase the bulk of IRAC colors for securely identified SMGs.  

The following describes the building of the IRAC color-color CDD for SMGs, starting with a brief discussion of the physical mechanisms responsible for the IRAC color-color properties of SMGs.

\subsubsection{The Physical Mechanism Behind the IRAC Color-Color Distribution of SMGs}
\label{sec:physics}

Due to the large redshift range at which SMGs have been detected thus far \citep[$z$$\sim$1-6 with a median redshift of $\sim$2-2.5; e.g.][]{sma04, cha05, are07, cha09, are11, yun12, smo12}, the IRAC bands probe a handful of spectral features over the rest-frame wavelength range of 0.5 to 4.0~$\mu$m.  In a typical star forming galaxy, this range includes the stellar photospheric feature at rest-frame 1.6~$\mu$m (the ``stellar bump''), which is the continuum emission from the old, low mass stars that form the bulk of the stellar mass of the galaxy, and, at wavelengths $\gtrsim$3~$\mu$m, emission from polycyclic aromatic hydrocarbons (PAHs).   In the case of extremely young galaxies, emission from OB stars can shift the bulk of the stellar emission to shorter wavelengths, depending on dust content.  Systems that are dominated by AGN show power-law emission from hot dust in this wavelength regime.

\begin{figure}
\centering
\includegraphics[scale=0.5, trim=10mm 0 2mm 2mm, clip]{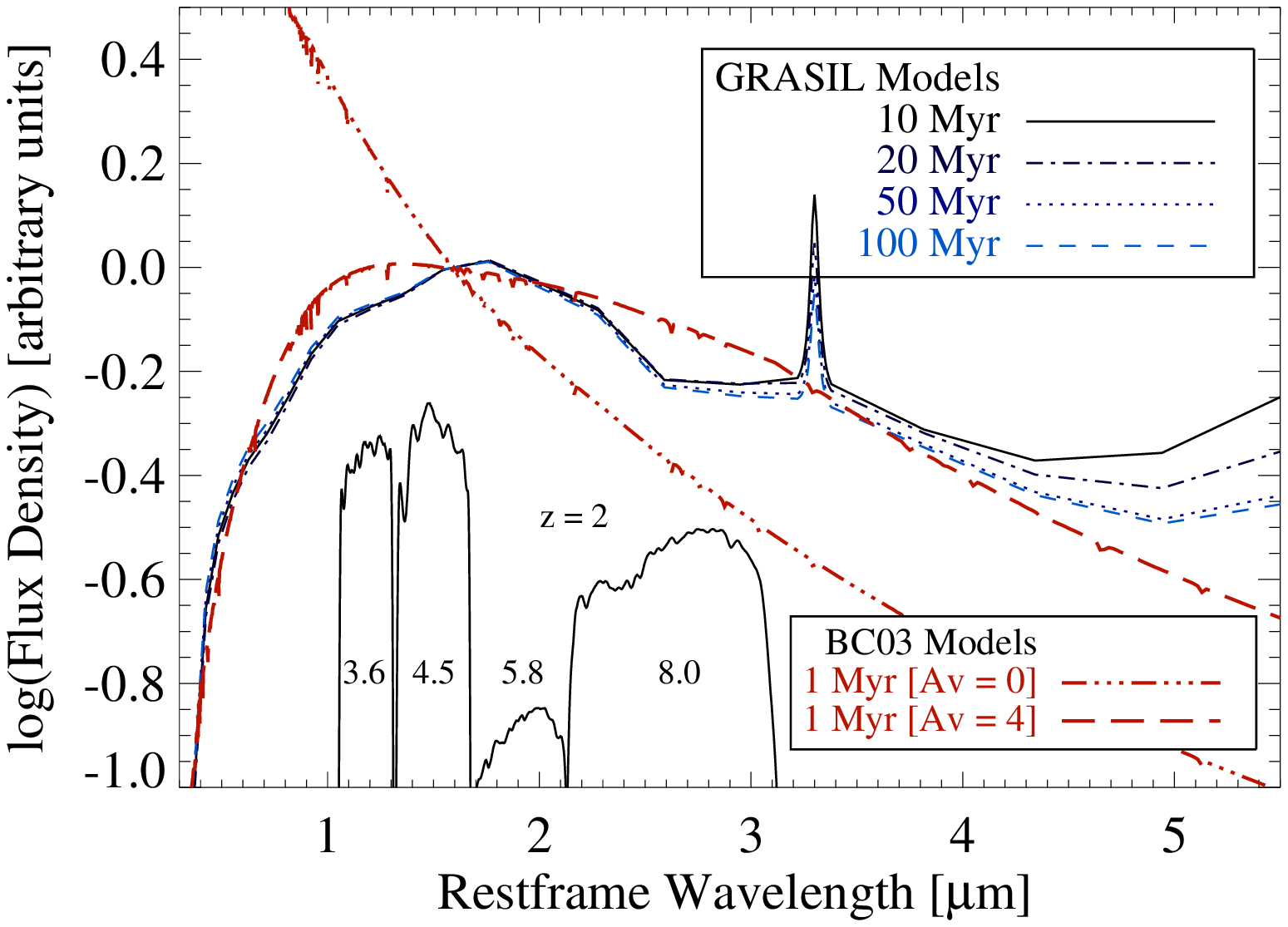} 
\includegraphics[scale=0.5, trim=10mm 0 2mm 2mm, clip]{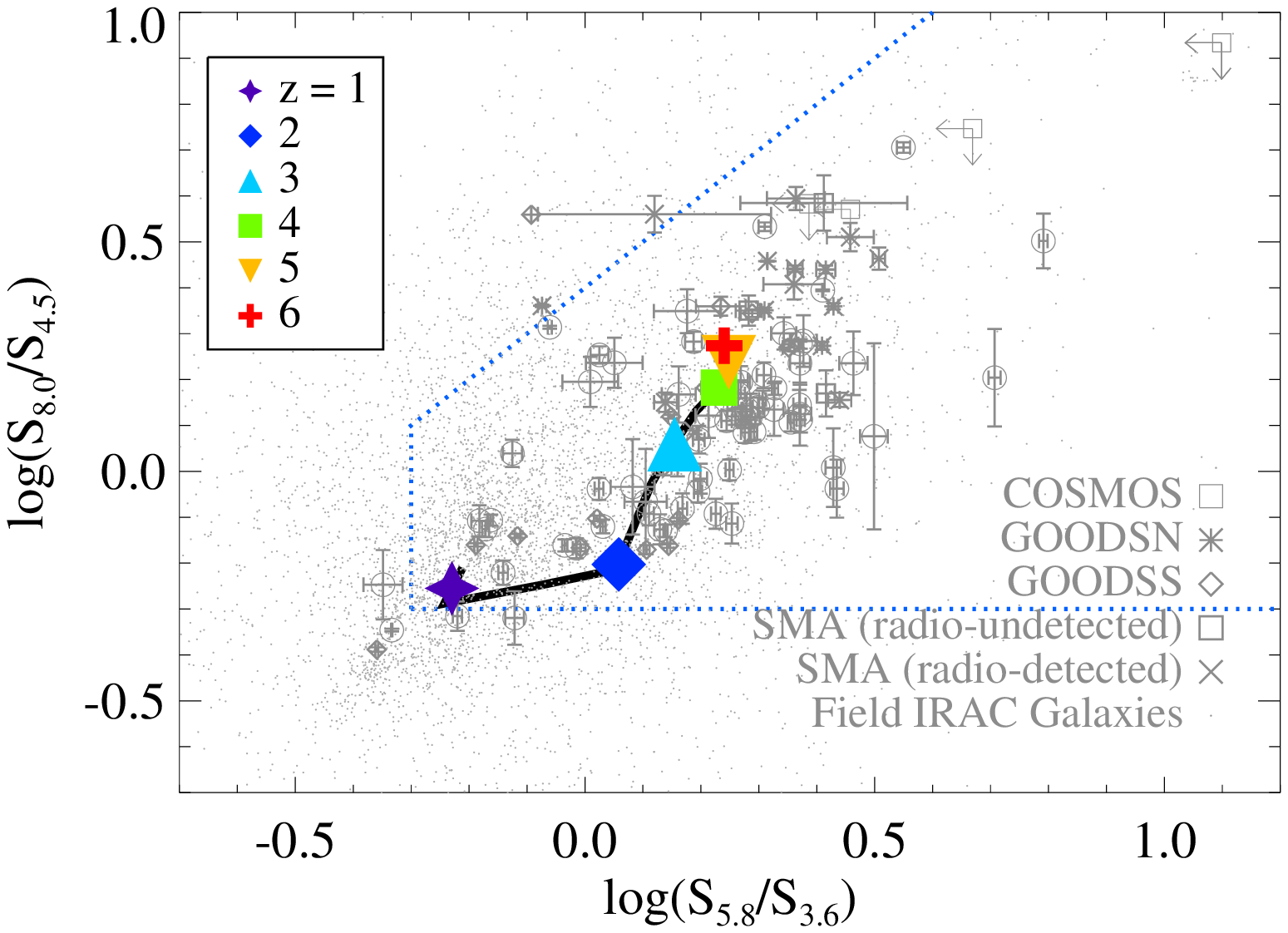}
\caption{\emph{(Top)} Starburst models of various ages with the IRAC filters at $z$=2 overlaid for reference.  The GRASIL models represent a typical SMG  with an underlying 1 Gyr old population and a starburst with ages of 10-100 Myr; these models show the ubiquitous stellar bump and a PAH feature at 3.4~$\mu$m.  The BC03 models represent an extremely young 1Myr starburst.  An extreme starburst with no dust (A$_v$ = 0) is not physical for SMGs, which are always dusty systems, so it is only shown for reference.  When dust is added (A$_v$ = 4), the emission from young stars is attenuated, forming a pseudo-stellar bump even in these extreme young starbursts.  In all models we examined, the stellar bump is evident by 10Myr.  \emph{(Bottom)} The redshifted IRAC color-color tracks of the GRASIL 50Myr SMG model between $z$=1-6 overlaid on the IRAC colors of the training set and a random subsample of IRAC field galaxies.}
\label{fig:models}
\end{figure}

To determine which of these features predominantly give rise to the observed IRAC colors of SMGs, we generated SMG galaxy models using the GRASIL SED modeling code \citep{sil98} based on the following properties:  SMGs are dominated by stellar emission as opposed to dust heated by AGN in the majority of cases \citep{ale05, pop08, men09, joh12}, the typical timescale of the SMG phase is $\sim$40-100 Myr \citep{gre05, nar10}, and submm galaxies are dusty systems \citep{bla02} which contain a pre-existing stellar population \citep{sma04,bor05,dye08,mic12}.  We create several GRASIL models with galaxy ages ranging from 1-2 Gyr, each undergoing an exponentially declining starburst with timescales ranging from 10-100 Myr and e-folding times ranging from 20-100 Myr.  Each model contains a moderate amount of dust with a dust to gas ratio of 5 per cent and an optical depth at 1~$\mu$m of 0.01.  For extremely young star forming systems, we compiled a set of models from \citet[BC03;][]{bru03} with ages 1-100 Myr, both with no extinction and with A$_V$ = 4 using the Calzetti extinction law \citep{cal01}. Figure~\ref{fig:models} (top) shows GRASIL models with a galaxy age of 1 Gyr with starburst ages of 10, 20, 50, and 100 Myr as well as two 1 Myr BC03 model SEDs, with and without dust extinction.  The BC03 1 Myr model without extinction is only for reference as all SMGs are dusty systems.  When extinction is added to extremely young systems, the emission from young stars is attenuated, creating a pseudo-stellar bump that is shifted to shorter wavelengths.  In both model sets, the stellar bump is prominent by $\sim$10 Myr, making it the dominant feature in the IRAC bands at the redshifts typical of SMGs.

%There are two features worth noting in addition to the stellar bump.  For SMGs at low redshift ($z$$lessim$1.5), the 3.2$\mu$m PAH feature is a potential contaminant that could cause a spike in the flux in one of the IRAC bands.  We expect, however, such low redshift sources to be rare.

Of course, the stellar bump is not unique to SMGs; previous studies have established that the IRAC bands are a useful redshift selection tool for star forming galaxies at $z$$>$1 due to the pervasiveness and robustness of the stellar bump and its shape \citep{pop06, far08, lon09, sor10}.  There is an additional complication in that at high redshift ($z$$>$4), the IRAC bands begin to sample only one side of the stellar bump, which can mimic a power-law shape, thus making it difficult to determine if these sources are star formation-dominated or AGN-dominated.  We argue that AGN are rare and it is already known that there is some overlap in SMG and AGN color in IRAC color-color space \citep{yun08,yun12}.  For further discussion on the IRAC colors of AGN, see \citet{don12}.  

The redshifted color-color tracks of our GRASIL models are shown in Figure~\ref{fig:models} (bottom) overlaid on the IRAC colors of our SMG training sample.  For comparison, the IRAC colors of a random sample of field galaxies are shown as well.  It can be seen from the plots in Figure~\ref{fig:models} that the separation in IRAC color-color space between SMGs and the bulk of field galaxies is predominantly a redshift effect.  As such, we are essentially using the IRAC colors to separate the high redshift SMG counterparts from the unrelated low redshift population. 

\subsubsection{Kernel Density Estimation}
\label{sec:kernel}

\begin{figure}
%\vspace*{-30pt}
%\hspace*{-20pt}
%\subfigure{\includegraphics[scale=0.5, trim=8mm 2mm 5mm 0, clip]{smeared_datapoints.eps}}
\vspace*{-10pt}
\hspace*{-20pt}
\subfigure{\includegraphics[scale=0.43, trim=8mm 0 5mm 5mm, clip]{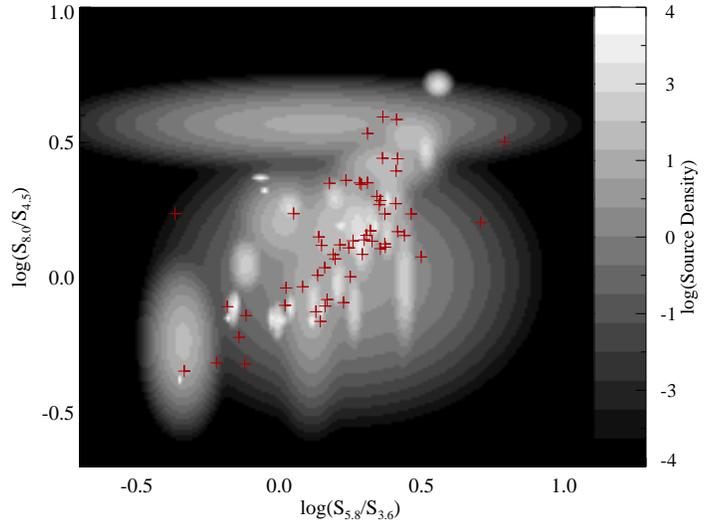}}
%\vspace*{-10pt}
%\hspace*{-20pt}
%\subfigure{\includegraphics[scale=0.5, trim=10mm 0 5mm 5mm, clip]{final_convolution.eps}}
\caption{An example of one cross-validation realization.  The white contours represent the ``working'' set, a randomly chosen half of the training set.  The datapoints in the working set have been replaced with normalized 2-d Gaussians in order to incorporate their photometric uncertainties.  The red pluses represent the ``cross-validation'' set.}

%As part of cross-validation, the working set is convolved with a kernel and then compared to the ``cross-validation'' set (the other half of the training set; red pluses) in order to find the kernel parameters which maximize the likelihood that these two sets are drawn from the same parent distribution. ***\emph{To commenters, do you find this figure illustrative or unneeded?}}
\label{fig:smeared}
\end{figure}

Our characteristic density distribution is developed using kernel density estimation, which provides a non-parametric estimator of the underlying probability density function of a data set \citep{par62, pis93, vio94, mer94}.  Functionally, this process convolves the IRAC colors of our training set with an optimized kernel to create a smoothed distribution which approximates the underlying parent distribution.  Cross-validation is used to optimized the kernel, with the optimal kernel width being a function of both the sample size of the data and the broadness (i.e. the standard deviation) of the intrinsic distribution. To build the CDD, we use the 102 radio- and SMA-identified IRAC counterparts of our training set described in Section~\ref{sec:irac}.  The steps for cross-validation are as follows:

\begin{enumerate}
\item Divide the training set in half randomly, forming a `working' set and a `cross-validation' set in IRAC color-color space.
\item  Incorporate the photometric uncertainties into the working set by replacing each datapoint within IRAC color space with a normalized 2-d Gaussian with a standard deviation defined by the 1$\sigma$ uncertainties on the colors.  This can be seen in Figure~\ref{fig:smeared}, overlaid with the datapoints that form the cross-validation set.
\item Next, select a kernel, which of its parameters to vary, and a suitable parameter space.  We choose a 2-d Gaussian kernel and allow its standard deviations, $\sigma_{\log\left(\frac{S_{5.8}}{S_{3.6}}\right)}$ and $\sigma_{\log\left(\frac{S_{8.0}}{S_{4.5}}\right)}$, to vary independently.  Preliminary tests showed that the optimized kernel has negligible rotation and so we do not allow the kernel to rotate in the following steps for computational simplicity.  Convolve the working set with the kernel while varying its parameters over the parameter space chosen.
\item For each set of parameters, compare the convolved working set to the cross-validation set.  
The optimized kernel parameters are those which maximize the likelihood that both the convolved working set and the cross-validation set are drawn from the same parent distribution.  This likelihood is calculated as 
\begin{equation}
L = \Pi_i C\left(\log\left(\frac{S_{5.8}}{S_{3.6}}\right)_i,  \log\left(\frac{S_{8.0}}{S_{4.5}}\right)_i, \sigma_{\log\left(\frac{S_{5.8}}{S_{3.6}}\right)}, \sigma_{\log\left(\frac{S_{8.0}}{S_{4.5}}\right)}\right)
\end{equation}
where $C\left(\log\left(\frac{S_{5.8}}{S_{3.6}}\right)_i, \log\left(\frac{S_{8.0}}{S_{4.5}}\right)_i, \sigma_{\log\left(\frac{S_{5.8}}{S_{3.6}}\right)}, \sigma_{\log\left(\frac{S_{8.0}}{S_{4.5}}\right)}\right)$ is the value of the convolved working set at the $i$th datapoint in the cross-validation set.
\item Finally, repeat steps i-iv to build a representative distribution of optimized kernel parameters.  We perform the cross-validation procedure 500 times and find that the distribution of optimized Gaussian widths does not significantly vary in the range of 0.07 to 0.21 for both colors, which indicates that the final kernel estimation results are not overly sensitive to the exact size of the Gaussian kernel.
\end{enumerate}

\begin{figure*}
{\includegraphics[scale=0.9, trim=10mm 0 5mm 0, clip]{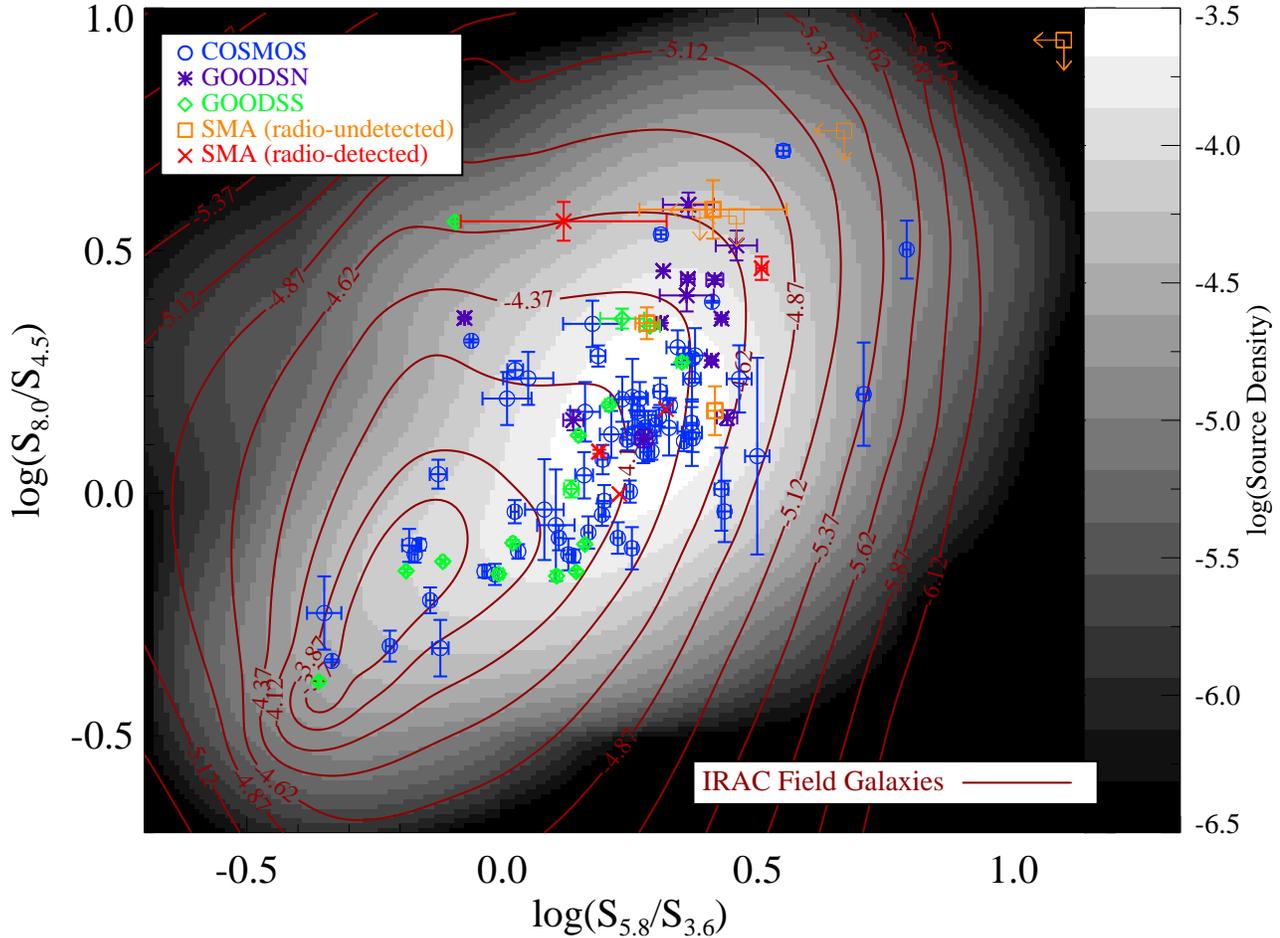}}
\caption{The final IRAC color-color characteristic density distribution for SMGs ($CDD_{SMG}$; white image contours) overlaid with the radio- and SMA-identified IRAC counterparts used to create it, as in Figure~\ref{fig:irac}.  Additionally, the CDD of all IRAC galaxies is shown ($CDD_{IRAC}$; red contours).}
\label{fig:convolution}
\end{figure*}

We adopt the mean values $\sigma_{\log\left(\frac{S_{5.8}}{S_{3.6}}\right)}$=0.14 and $\sigma_{\log\left(\frac{S_{8.0}}{S_{4.5}}\right)}$=0.15 for the final convolution of the 102 IRAC counterparts in the training set with an optimized kernel.  Our final IRAC color-color $CDD_{SMG}$\footnote{Available at \url{www.astro.umass.edu/aztec/CounterpartID/counterpartid.html}} for submillimetre galaxies can be seen in greyscale in Figure~\ref{fig:convolution} overlaid with the original IRAC datapoints used to create it.  

\begin{figure}
\vspace*{-10pt}
\hspace*{-25pt}
{\includegraphics[scale=0.36, trim=6mm 0 0 10mm, clip]{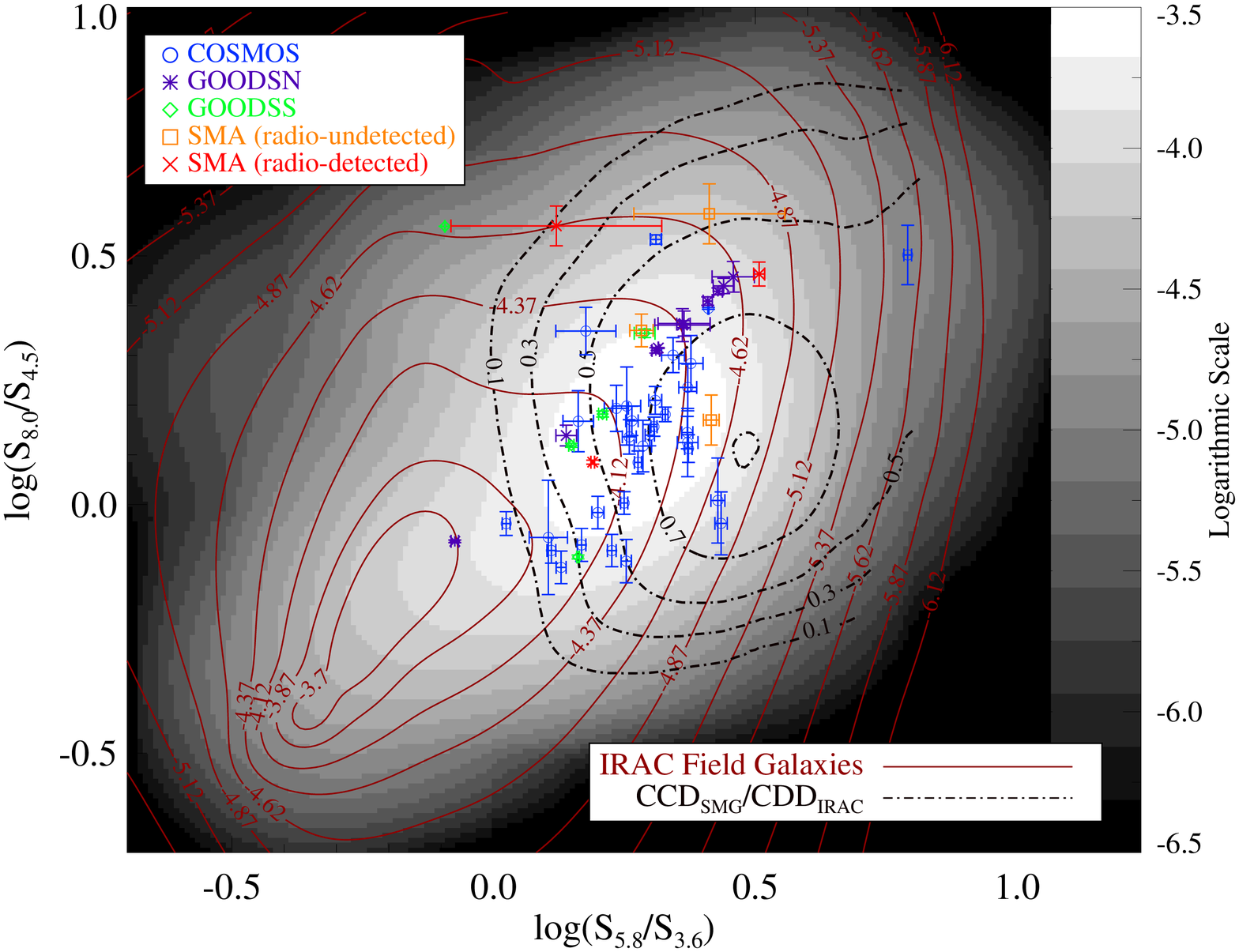}}
\caption{The ratio $CDD_{SMG}$/$CDD_{IRAC}$ in black, dash-dot contours, overlaid on $CDD_{SMG}$ in the white image contours, and $CDD_{IRAC}$ in red, solid contours.  The datapoint shown here are the IRAC counterparts in the training set which were identified correctly by their likelihood ratio.}
\label{fig:cratio}
\end{figure}

To use this characteristic density distribution to identify counterparts to the rest of our SMGs, we must also determine the IRAC color distribution for the full IRAC population.  We use the same kernel density estimation discussed above on all IRAC galaxies that are detected in all four IRAC bands across the three fields, $\sim$295,000 sources, and find a smaller range in optimal kernel widths, 0.02-0.06.  We adopt the mean value of 0.04 for both standard deviations of the Gaussian kernel.  The final distribution $CDD_{IRAC}$, shown in red contours in Figure~\ref{fig:convolution}, peaks in a bluer regime of IRAC color-color space than the distribution for SMGs.  This is expected since the full IRAC sample is dominated by $z$$\lesssim$1 galaxies.  At these lower redshifts, the IRAC bands sample the more complex PAH emission features, and the variety in observed galaxy SEDs at low redshifts accounts for the large scatter in IRAC color-color space of the general IRAC population.  Regardless, the distribution of the field IRAC galaxies is distinct from that of the SMG population, which allows us to make use of this method for counterpart identification.

\section{Identifying IRAC Counterparts to SMGs}
\label{sec:counter}

\subsection{The Likelihood Ratio and Reliability Factor}
\label{sec:rank}

The relative probability that an IRAC galaxy is the counterpart of an SMG as opposed to being an unrelated foreground/background source is given by the likelihood ratio \citep{sut92}:

\begin{equation}
\label{eqn:likeratio}
\lambda = \frac{CDD_{SMG}\left(\log{\left(\frac{S_{5.8}}{S_{3.6}}\right)}, \log{\left(\frac{S_{8.0}}{S_{4.5}}\right)}\right)}{CDD_{IRAC}\left(\log{\left(\frac{S_{5.8}}{S_{3.6}}\right)}, \log{\left(\frac{S_{8.0}}{S_{4.5}}\right)}\right)} P(D;S/N)
%\lambda = \frac{\mbox{(Q(\log(S}_{5.8}/\mbox{S}_{3.6}),\log(\mbox{S}_{8.0}/\mbox{S}_{4.5}))\mbox{P(D;S/N)}}{\mbox{N(S}_{5.8}\mbox{/S}_{3.6},\mbox{S}_{8.0}\mbox{/S}_{4.5})}
\end{equation}
where $CDD_{SMG}\left(\log\left(\frac{S_{5.8}}{S_{3.6}}\right), \log{\left(\frac{S_{8.0}}{S_{4.5}}\right)}\right)$ is the value of the CDD of SMGs at the location of a potential IRAC counterpart in IRAC color-color space, $CDD_{IRAC}\left(\log\left(\frac{S_{5.8}}{S_{3.6}}\right), \log{\left(\frac{S_{8.0}}{S_{4.5}}\right)}\right)$ is the value of the CDD of the general IRAC population for that potential counterpart, and $P(D;S/N)$ is given by Eqn.~\ref{eqn:pu} integrated over $D$-$\epsilon$ to $D$+$\epsilon$, where $D$ is the distance of a given IRAC galaxy from the centre of the detected submm emission and $\epsilon$=0.2$^{\prime\prime}$ is the positional accuracy of the IRAC observations.  $P(D;S/N)$ represents the probability that the true SMG position is found at distance $D$, based on the S/N and FWHM of the submm detection.

%is the likelihood that an IRAC galaxy is the counterpart based on the IRAC color-color characteristic density distribution of SMGs, $CDD_{IRAC}\left(\log\left(\frac{S_{5.8}}{S_{3.6}}\right), \log{\left(\frac{S_{8.0}}{S_{4.5}}\right)}\right)$ is the likelihood that an IRAC galaxy is unrelated to the SMG and is given by the IRAC color-color distribution for the field IRAC galaxies, and  $P(D;S/N)$ is given by Eqn.~\ref{eqn:pu} integrated over $D$-$\epsilon$ to $D$+$\epsilon$, where $D$ is the distance of a given IRAC galaxy from the centre of the detected submm emission and $\epsilon$=0.2$^{\prime\prime}$ is the positional accuracy of the IRAC observations.  $P(D;S/N)$ represents the probability that the true SMG position is found at distance D, based on the S/N and FWHM of the submm detection.

The likelihood ratio provides a means of ranking IRAC galaxies in the proximity of SMGs based on their colors and positions.  Since there is no IRAC color combinations that is completely unique to SMGs, this ranking is relative within a set of potential counterparts.  To determine the statistical significance of any given likelihood ratio, we calculate its 1$\sigma$ uncertainty by including the measured errors in the IRAC colors.  The ratio $CDD_{SMG}$/$CDD_{IRAC}$ can be seen in Figure~\ref{fig:cratio} in black, dash-dot contours, overlaid on $CDD_{SMG}$ in greyscale and $CDD_{IRAC}$ in red, solid contours.  Due to the rapid falling off of both the SMG and IRAC CDDs in IRAC color-color space, a floor value of $\log{f}$ = -5.7 was imposed on both distributions when plotting the ratio to suppress noise features in areas with sparse data.  This floor was not used in the calculation of the likelihood ratio and we have verified that these noise features are not affecting our counterpart identification results.  The peak of the $CDD_{SMG}$/$CDD_{IRAC}$ contours show the area in which the color prior portion of the likelihood ratio is at a maximum.

Additionally, we adopt the reliability factor, as described in \citet{sut92}, which provides a single statistic for a given potential counterpart which incorporates its individual LR, the LRs of all other potential counterparts around that SMG, and the probability that the true counterpart is detected. The reliability factor for potential IRAC counterpart $j$ takes the form
\begin{equation}
\label{eqn:rel}
R_j = \frac{\lambda_j}{\sum_i \lambda_i + (1-X)}
\end{equation}
where $\sum_i$ $\lambda_i$ is the sum over the likelihood ratios of all potential IRAC counterparts around a given SMG and $X$ is the probability that the true IRAC counterpart is included in the four-band IRAC catalog we are using.  Though interferometric studies indicate a high probability that the majority of SMGs are detected at $\geq$1~$\mu$Jy at 3.6 and 4.5~$\mu$m \citep[e.g.][]{you07,you09, hat10, ika11}, our technique requires detections at 5.8 and 8.0~$\mu$m as well.  These bands are less sensitive than the two shorter IRAC channels, though our training set suggests a high percentage (99$\%$) of counterparts are detected in all bands.  The overall fraction of IRAC sources not detected in all bands therefore provides an estimate of $X$.  

To control for possible biases in the properties of IRAC galaxies that are nearby SMGs as opposed to the general field population, we elect to estimate $X$ only from IRAC galaxies within the vicinity of SMGs, including both our training sample and the radio-undetected SMGs.  To do this, we identify all IRAC sources within the search radii described in Section~\ref{sec:posun} and listed in Tables~\ref{app:robust}-\ref{app:main}.   We find that the fraction of IRAC sources detected in all four bands is 0.83, 0.78, and 0.73 for GOODS-N, GOODS-S, and COSMOS, respectively, and we use these as approximations for $X$ in Eqn~\ref{eqn:rel}.  We note that this fraction only changes by $<$10 per cent if we include all IRAC catalog sources, suggesting no strong biases to this fraction from redshift selection effects.

In principle, $\sum_i$ $\lambda_i$ for a given SMG in Eqn.~\ref{eqn:rel} runs over the entire IRAC four-band catalog.  In practice, however, $\lambda_i$ decreases rapidly with increasing distance between the mm peak emission and the IRAC position (Eqn~\ref{eqn:likeratio}).  For this reason, we limit the sum in the denominator in Eqn~\ref{eqn:rel} to IRAC galaxies within the search radii for the SMG, defined where $P($$>$$D;S/N)$ = 0.01, to simplify computations.

\begin{table*}
\begin{minipage}[!ht]{0.6\linewidth}
\caption{Statistical results of counterpart identification.  Columns 2-5 show the distribution of ranks for the training set per field.  Being ranked first is considered a correct identification.  Additionally shown are the number of robust and tentative reliabilities in the training set and the radio-undetected set.}
\label{tbl:rank}
\begin{tabular}{lcccccc}
\hline
Field & Ranked & Ranked & Ranked & Ranked & \multicolumn{2}{c}{$R$$\ge$0.8 ($R$$\ge$0.6)} \\
& 1st & 2nd & 3rd & 3rd+ & Training Set\tablenotemark{a} & Radio-Undetected \\
\hline
GOODS-N & 11 (85$\%$) & 2 (15$\%$) & 0 & 0 & 6/6 (9/9) & 2 (8)   \\
GOODS-S & 5 (33$\%$) & 5 (33$\%$) & 2 (13$\%$) & 3 (20$\%$) & 2/3 (3/7) & 6 (14)  \\
COSMOS & 33 (49$\%$) & 15 (22$\%$) & 8 (11$\%$) & 12 (17$\%$) & 11/12 (18/28) & 21 (48) \\
\hline
\end{tabular}
{\scriptsize $^a$The number of correctly identified counterparts in the training set with a reliability factor above 0.8 (0.6)
compared to the total number of IRAC galaxies assigned a reliability above that value.}
\end{minipage}
\end{table*}

\subsection{Counterpart Identification via the Reliability Factor}
\label{sec:analysis}

\subsubsection{Verification of the Method: the Training Set}

We evaluate our technique with a blind-test against our radio- and SMA-identified training set.  All IRAC galaxies within the appropriate search radius of each SMG are ranked using the likelihood ratio (Eqn.~\ref{eqn:likeratio}) and assigned a reliability (Eqn.~\ref{eqn:rel}).  The likelihood ratio, its 1$\sigma$ errors, and reliability of the radio- or SMA-identified counterpart as well as additional top ranked potential IRAC counterparts for the training set can be seen in Table~\ref{app:robust}.  

In GOODS-N, 11/13 (85$\%$) IRAC counterparts are identified correctly, i.e. ranked first among the other possibilities using the likelihood ratio.  Three of these sources (AzTEC/GN3, AzTEC/G5, and AzTEC/GN14) are confirmed by interferometry \citep{bar12}.  In GOODS-S, 5/15 (33$\%$), and in COSMOS, 33/68 (49$\%$), are identified correctly, including one radio-undetected SMG (AzTEC/C4), one multiple radio system (AzTEC/C42, see below), and two radio-detected SMGs (AzTEC/C18 and AzTEC/C22) which were imaged with the SMA.  The correctly identified counterparts, as well as SMA-confirmed counterparts \citep{you07, you09}, can be seen in Figure~\ref{fig:cratio}, overlaid on contours showing $CDD_{SMG}$, $CDD_{IRAC}$, and the ratio $CDD_{SMG}/CDD_{IRAC}$.  The difference in the success rates of GOODS-N versus COSMOS and GOODS-S stems from the difference in beamsizes:  18$^{\prime\prime}$ in GOODS-N compared to 33$^{\prime\prime}$ and 30$^{\prime\prime}$ in COSMOS and GOODS-S.  Possible explanations for the more subtle differences between COSMOS and GOODS-S are discussed in Section~\ref{sec:aztecflux}. The distribution of counterpart rankings (1st, 2nd, 3rd, and 3rd+) in each field can be seen in Table~\ref{tbl:rank}.  

%Two SMA observations \citep[SMA7 and SMA10;][]{you07,you09} did not match any AzTEC sources in the ASTE catalog; these sources were used in the development of the IRAC color-color distribution, but they were not part of the blind test.  For two additional SMA observations (AzTEC/C2 and AzTEC/C10) used in determining the IRAC color-color distribution, our analysis did not select the correct IRAC counterpart.

\begin{figure}
\includegraphics[scale=0.55, trim=10mm 0 0 0, clip]{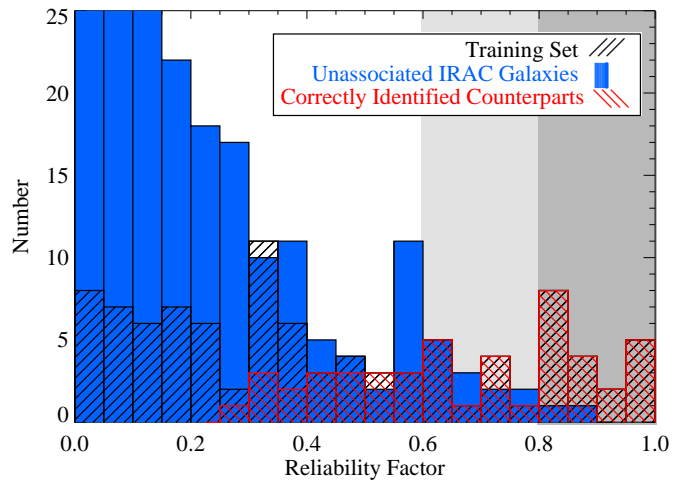}
\caption{The distribution of reliabilities for the IRAC candidates in the radio- and SMA-identified training set.  The entire training set (black) is correctly identified (ranked first via the likelihood ratio; red) above $R$=0.5, though there is contamination from possibly unassociated IRAC galaxies at almost all reliabilities (solid blue).  The darker shaded region represents what we consider to be the robust range of reliabilities, while the lighter shaded region defines a tentative range.}
\label{fig:rel}
\end{figure}

The distribution of reliabilities for the radio- and SMA-identified training set can be seen in Figure~\ref{fig:rel}.  A large $R$ value indicates a high probability that a given IRAC galaxy is the true counterpart as well as a high probability that only a single galaxy is responsible for the submm/mm emission.  A low $R$ value indicates a low probability of being a counterpart and/or that the corresponding SMG consists of multiple galaxies blended within the beam (see Section~\ref{sec:blends}).  
%Previous studies have determined that roughly 10-20 per cent of SMGs are blends of multiple sources which are truly associated with more than one radio source \citep{pop06,ivi07,cle08,cha09,yun12,smo12}. 
Figure~\ref{fig:rel} shows that for $R$$\ge$0.5, all radio- or SMA-identified IRAC counterparts are identified correctly, whereas for $R$$\le$0.5, many radio-identified IRAC counterparts are not. We suggest a conservative cut-off in the reliability factor would be at $R$$\ge$0.8, where there are only two misidentifications out of nineteen.  This cut-off is the same as that found in similar studies utilizing the reliability factor \citep[e.g.][]{smi11}.  Additionally, an analysis of the field-by-field reliability factors suggests that extending the cutoff to $R$$\ge$0.6 is appropriate in fields analogous to GOODS-N, where 9/9 (100$\%$) IRAC galaxies assigned $R$$\ge$0.6 are the correct counterpart.  The range 0.6$\le R\le$0.8 is more appropriately named a `tentative' counterpart list in COSMOS with only 18/28 (64$\%$) IRAC galaxies assigned $R$$\ge$0.6 being the correct counterpart.  In GOODS-S, this ratio drops to 3/7 (43$\%$) (see Table~\ref{tbl:rank}). %Alternative interpretations of the reliability factor and its use in the face of possible blending as well as other methods for establishing the robustness of LRs exist in the literature  \citep[e.g.][]{cha11}.

\subsubsection{Multiple Radio Systems and Radio Undetected SMGs}

The likelihood ratios and reliabilities for SMGs with multiple radio counterparts can be seen in Table~\ref{app:multiples}.  Some of these systems may be blended, multiple systems, and, as such, the IRAC counterparts for many of these systems have low reliabilities.  There are a small number of cases (5/27), however, where a single IRAC candidate has an $R$$\ge$0.8 and is thus likely associated with only one or none of the radio sources.  One of these cases, AzTEC/C42, was observed with the SMA, which confirmed that the submm emission is only coming from one of the two radio sources within the AzTEC beam.  The IRAC galaxy associated with the true radio counterpart was correctly identified with our technique.
%, with R=0.88.  

%\subsubsection{Identifying IRAC Counterparts to Radio-Undetected SMGs and Comparison to Previous Studies}

The ranking and reliabilities for the IRAC galaxies corresponding to the radio-undetected SMG set\footnote{A list of AzTEC sources and potential IRAC counterparts, including photometry, is available at \url{www.astro.umass.edu/aztec/CounterpartID/counterpartid.html}.} can be seen in Table~\ref{app:main}.  There are 29 (70) radio-undetected SMGs with $R$$\ge$0.8 ($R$$\ge$0.6)  (see Table~\ref{tbl:rank}).

\section{Discussion}
\label{sec:disc}

\subsection{Source Blending}
\label{sec:blends}

Multiple studies have established that some subset of SMGs are in fact blends of multiple sources of submm emission which appear as one source in the large beams of single-dish submm observations.  Our best estimates of the blending fraction come from interferometric studies, which have determined that 10-30$\%$ of known SMGs are multiple systems \citep{you07, you09, smo12, bar12}.  A recent study of LABOCA sources imaged with ALMA determined that the overall blending fraction in their sample of 88 is 26$\%$, though this fraction is higher for the very brightest SMGs \citep[$S_{870\mu m}\ge$9 mJy;][]{kar12}.  

The formulation of the reliability factor is such that it assumes there is only one unique counterpart.  In the event of blending, the reliability of a potential counterpart will reflect that there are multiple sources with a large LR values by assigning these sources correspondingly low reliabilities.  These cases are indistinguishable from those in which an unrelated IRAC galaxy is assigned a high LR due to a chance superposition and a redshift that is consistent with the peak in the IRAC color-color distribution of SMGs.  For this reason, the reliability factor is biased against the identification of blended submm systems.

Currently, there is no formulation of the reliability factor that accounts for the possibility of multiple counterparts.  An alternative to the reliability factor was presented in \citet{cha11}, which used simulations to determine a cut-off threshold on the likelihood ratio above which the rate of false positives was $\sim$10 per cent.  We performed a preliminary analysis to determine the utility of this technique with our dataset as follows:  in COSMOS and GOODS-N, we choose 10,000 random locations with the areas around known SMGs masked out and calculate the LR of all IRAC galaxies within a search radius as described in Section~\ref{sec:id}.  As the LR incorporates positional information based on the S/N of the submm observation, we repeated this process for a range of S/N bins.  A likelihood ratio threshold is determined such that 90 per cent of the IRAC galaxies are rejected, which corresponds to a 10 per cent false positive rate.  The LR of each potential IRAC counterpart is then divided by the appropriate threshold according to the S/N of the corresponding SMG. As in \citet{cha11}, a normalized likelihood ratio (nLR) of greater than one indicates a candidate counterpart.

An analysis of the results of using a nLR finds a higher fraction of SMGs with multiple sources assigned nLR$>$1, indicating blends, than we would expect from interferometric studies.  In COSMOS, we find that 31/68 (46$\%$) of SMGs have more than one counterpart with nLR$>$1 and in GOODS-N this number is 8/13 (62$\%$).  In addition, we examined multiple AzTEC sources that correspond to SCUBA sources in \citet{bar12}, which were determined to not be blended at the resolution of the SMA.  All five sources have multiple counterparts with nLR$>$1.

These results indicate that the normalized likelihood ratio, when used in conjunction with our dataset and methods, suffers from contamination from unrelated IRAC galaxies. An additional difficulty with these simulations arises in small fields such as GOODS-N, where there are only $\sim$100 unique locations in the map given our typical search radius.  While the normalized likelihood ratio is not biased against blended systems, it presents cases where blending is indistinguishable from confusion due to unrelated IRAC galaxies in excess of the 10$\%$ false positive rate expected.  As such, we find the reliability factor to be more conservative in identifying unambiguous, single counterparts and do not present the normalized likelihood ratios in this work.

\subsection{Dependence on the 1.1 mm and IRAC Observations}
\label{sec:aztecflux}

In the previous section, we found that the accuracy of our technique, as blind-tested by our training set, improves with decreasing beamsize as we would expect.  Here we explore additional effects on this technique; particularly we look at differences between the COSMOS and GOODS-S fields, since, despite being observed with similar beamsizes, our technique performs more accurately in COSMOS over GOODS-S by $\sim$10 per cent.  

\begin{figure}
\includegraphics[scale=0.55, trim=12mm 0 2mm 2mm, clip]{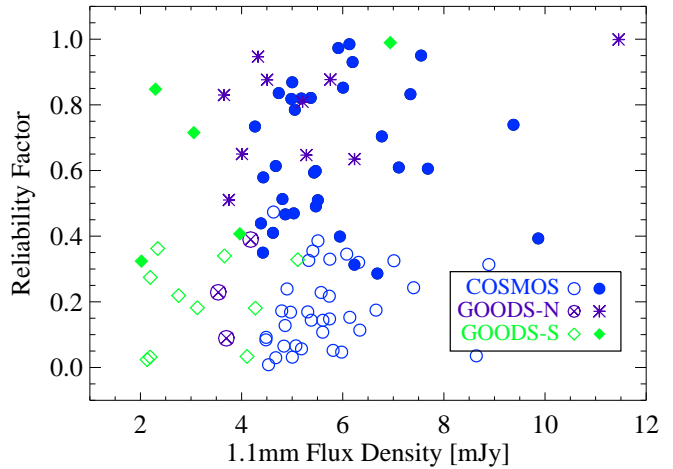}
\caption{The dependence of the reliability factor on the AzTEC flux density for training set counterparts that are 1) ranked first (identified correctly) with our technique (filled symbols) and 2) those that are not ranked first (open symbols).  There is no correlation, which suggests that the AzTEC flux density does not directly affect our counterpart identification technique.}
\label{fig:vs}
\end{figure}

The submm number counts reveal that SMGs in GOODS-S are systematically fainter than in COSMOS \citep{sco12}.  Since our technique depends on the detectability of SMGs in all IRAC bands, we consider whether fainter submm sources implicitly translate to fainter IRAC counterparts.  Our training set indicates that this is not the case as we find no correlation between the 1.1 mm flux density and any of the IRAC bands.  This is what we would expect, given that the 1.1 mm flux density is powered by dust emission and the IRAC bands primarily probe the mostly dust-free stellar bump at redshifts characteristic of SMGs.    Additionally, we considered whether brighter or fainter SMGs have more easily identifiable counterparts due to some external factor which affects their IRAC colors, such as their redshift distribution.  Figure~\ref{fig:vs} shows the reliabilities of SMGs in our training set versus their 1.1 mm flux density; the two variables have a Pearson correlation coefficient of $\sim$0.05, consistent with being uncorrelated.  Cutting the GOODS-S AzTEC catalog to the depth of the COSMOS AzTEC catalog similarly produces no improvement in the counterpart identification results in GOODS-S.

Next, we consider what effect the depth of the IRAC catalog may have on our analysis.  %In Section~\ref{sec:posun}, we found that 96/107 (90$\%$) of radio counterparts are also detected in all four IRAC bands.  If we break down these numbers by field, we find that 13/18 (72$\%$), 15/15 (100$\%$), and 68/74 (92$\%$) of radio counterparts are detected with IRAC for GOODS-N, GOODS-S, and COSMOS, respectively.  GOODS-N contains the deepest IRAC catalog of the three fields, indicating that the depth of the IRAC catalogs is not driving the detectability of the IRAC counterparts in this work.  
To directly compare GOODS-S and COSMOS, we cut the GOODS-S IRAC catalog to the depth of the COSMOS catalog, removing any sources with a 3.6~$\mu$m detection less than 1~$\mu$Jy.  This results in an improved identification rate, with 7/15 (46$\%$) IRAC counterparts to SMGs in GOODS-S correctly identified.  This is similar to the 49 per cent identification rate in COSMOS.  This effect is due to the large beamsize of the AzTEC observations, necessitating a larger search radius for counterparts, and the depth of the IRAC observations in GOODS-S increasing the chances of detecting a high redshift galaxy which is unrelated to the SMG, but, due to its redshift, has similar IRAC colors. These chance associations can be mitigated by higher resolution submm/mm observations or by cutting the IRAC catalog, as SMGs will be brighter in the IRAC bands than other typical high redshift galaxies due to having larger stellar masses by about an order of magnitude \citep[see][and references therein]{hai11}.

\subsection{Dependence on the Radio Flux Density}
\label{sec:multiples}

The utility of our technique depends in large part on the applicability of the IRAC CDD to radio-undetected SMGs.  We test this in two ways: first, we check that our IRAC color distribution is robust against the depth of the radio observations.  Second, we split our training set into bright and faint radio counterparts, to determine if there is a systematic difference in their IRAC colors.

The effects of the radio depth on our technique is tested by extending the COSMOS radio catalog from the original 4.5$\sigma$ cutoff to 3$\sigma$ \citep[e.g.][]{cha05, yun12}.  Since the false positive rate increases dramatically with decreasing S/N, we identify all $\ge$3$\sigma$ peaks above the local noise in the radio map and then match these peaks to the COSMOS IRAC catalog.  Only those radio peaks with a corresponding IRAC counterpart within 2$^{\prime\prime}$ are considered.

\begin{figure}
\includegraphics[scale=0.55, trim=12mm 0 2mm 2mm, clip]{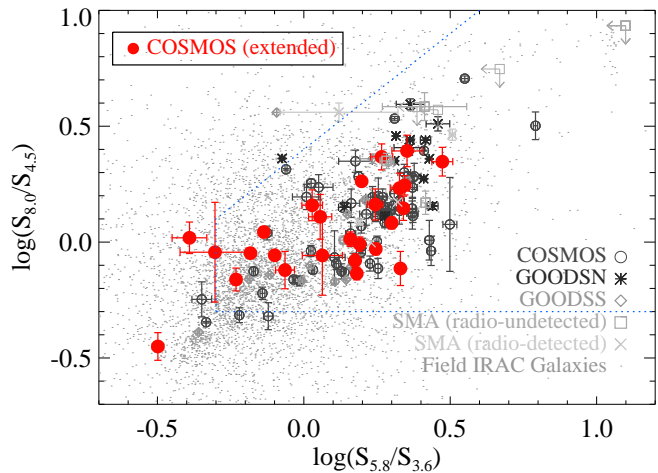}
\caption{IRAC colors of COSMOS radio sources in the extended catalog (S/N=3-4.5; red filled circles) overlaid on the original training set used in this study (see Figure~\ref{fig:irac}).}
\label{fig:extended}
\end{figure}

Originally, we found 74 single radio counterparts to SMGs, plus 15 multiple radio systems.  Using the deeper radio catalog, we find 26 new single radio detections and 3 new multiple radio detections around previously radio-undetected SMGs.  In addition, 7/74 COSMOS SMGs in the original training set were found to have a second radio source within their search radius and one double system was found to have a third radio source (AzTEC/C184).  This makes a total of 118/189 (62$\%$) radio-detected SMGs in COSMOS, including 25 (13$\%$) multiple radio systems (Tables~\ref{app:robust}-\ref{app:multiples}). The IRAC colors of the new single radio detections can be seen overlaid on our original IRAC counterpart set in Figure~\ref{fig:extended}. A 2-D Kolmogorov-Smirnov (KS) Test \citep{fas87} returns a value of 0.84, indicating the original training set and extended radio counterparts are consistent with being drawn from the same parent distribution. When analysed using the IRAC CDD built with the original training set, 12/26 (46$\%$) of the new radio sources were ranked first (4 with $R$$\ge$0.8).

Combining the new single radio counterparts with the previous training set (and removing the new multiple radio systems), we perform the same kernel density estimation as before and apply this new IRAC color-color CDD in a blind test as in Section~\ref{sec:rank}.  We find that the addition of 26 new radio counterparts has a negligible effect on the resulting likelihood ratios and reliabilities assigned to each potential IRAC counterpart. This indicates that our IRAC color-color CDD is well-sampled and robust against the depth of the radio observations.

\begin{table}
\begin{minipage}[!ht]{0.6\linewidth}
\caption{Summary of Cryogenic Mission $Spitzer$ Legacy Programs}
\label{tbl:spitzer}
\begin{tabular}{lcc}
\hline
Survey & Area & Reference \\
\hline
GOODS & 280 arcmin$^2$ & \citet{dic00} \\
SIMPLE & $\sim$0.5 deg$^2$ & \citet{dam11} \\
SpUDS & 1 deg$^2$ & \citet{dun07} \\
S-COSMOS & 2 deg$^2$ & \citet{san07} \\
SDWFS & 10 deg$^2$ & \citet{ash09} \\
SWIRE & $\sim$60 deg$^2$ & \citet{lon03} \\ 
\hline
\end{tabular}
\end{minipage}
\end{table}

As an additional check against potential bias on the IRAC colors used in this technique, we split our training set in half to analyse the IRAC colors of radio counterparts with radio flux densities greater than and less than 65~$\mu$Jy.  A 2-D KS test of the two resulting distributions in IRAC color-color space gives a value of 0.98, confirming that the IRAC colors of SMGs with radio flux densities above and below 65~$\mu$Jy are consistent with being drawn from the same parent distribution. This indicates that there is no dependence on the IRAC characteristic density distribution on the radio flux density.

Given these two tests, we see no evidence that the IRAC colors of bright and faint radio counterparts are systematically different, which is consistent with the detectability of radio counterparts being a function of the depth of the radio observations and a selection effect due to the strong positive k-correction in the radio regime.  This result gives us confidence that the IRAC color-color CDD developed in this study can be applied to our radio-undetected SMG sample.
%\begin{figure}
%\hspace{-20pt}
%\subfigure{\includegraphics[scale=0.45, trim=20mm 0 0 0, clip]{final_convolution_gt_65.eps}}
%\hspace{-20pt}
%\subfigure{\includegraphics[scale=0.45, trim=20mm 0 0 5mm, clip]{final_convolution_lt_65.eps}}
%\caption{The IRAC counterparts to radio galaxies with flux greater than 65~$\mu$Jy (top) and less than 65~$\mu$Jy (bottom) convolved with a gaussian kernel to form IRAC color-color characteristic density distributions. }
%\label{fig:radiocut}
%\end{figure}

%\subsection{Comparison with Previous Studies}

%\emph{What to add here??}

\subsection{In the Era of Warm $Spitzer$, WISE, and JWST}
\label{wise}

In this study, we have utilized data from the four IRAC bands available during $Spitzer's$ cryogenic mission.  This mission surveyed 10s of square degrees on the sky over several wide, deep Legacy Programs, a summary of which can be seen in Table~\ref{tbl:spitzer}. These fields will be natural targets for future submm and mm observations.

In May 2009, the cryogenic mission of $Spitzer$ ended and it began the ``warm'' mission phase which includes only the 3.6 and 4.5~$\mu$m channels.    Unfortunately, the close spacing of the two warm IRAC channels means that at $z$$\gtrsim$2, both bands fall on the same side of the stellar bump and their redshift discriminating power is much diminished.  As such, performing a characteristic density distribution analysis as seen here is not beneficial for SMGs discovered in fields with only warm $Spitzer$ coverage, and we must resort to simpler selection methods, such as adopting a straight color cut or using a color cut in the formation of an IRAC color prior or P-statistic \citep{yun08, yun12, cha11}. 

The Wide-field Infrared Survey Explorer \citep[WISE;][]{wri10} is an infrared survey which will map the entire sky at 3.4, 4.6, 12, and 22~$\mu$m to 5$\sigma$ depths of 0.08, 0.11, 1, and 6 mJy.  Though this is far too shallow to detect the infrared counterparts of normal SMGs, WISE will be able to detect strongly lensed submm background galaxies, with magnification factors up to 30x \citep{ega10}.  Though these lensed sources are rare \citep[e.g.][]{swi10,neg10,con11}, surveys with $Herschel$ \citep[e.g.][]{ega10, eal10, oli10}, the SPT \citep{vie10}, and ACT \citep{mar11} and future large-scale SMG surveys with the Large Millimeter Telescope (LMT) and SCUBA-2 will dramatically increase the number of known lensed submm galaxies.  The three short WISE bands will bracket the stellar bump at $z$$>$2 and have counterpart discriminating power similar to that derived in this study.

Hubble's successor, the James Webb Space Telescope (JWST), will have two instruments, NIRCam and MIRI, which together will provide imaging in the range 0.6 to 28~$\mu$m, covering the wavelength regime of $Spitzer$ IRAC and MIPS 24~$\mu$m.  With sensitivities $\sim$50x better than IRAC, JWST will be able to detect mid-infrared counterparts out to very high redshifts in multiple bands which bracket the stellar bump.

%One such lensed SMG have already been detected as part of the Herschel HerMES survey.  Confirmed at $z$=2.96, the SMG is undetected at 3.6$\mu$m, but detected at 0.4mJy at 4.5$\mu$ and 0.6 mJy at 8.0$\mu$m.  

\subsection{Use in Ongoing and Future Submm/mm Surveys}
\label{lmtsim}

We are currently seeing major advances in single-dish submm/mm observations, with improved resolution for observations at 450-1100~$\mu$m.  Large surveys are being carried out by SCUBA-2 on the JCMT \citep[450, 850~$\mu$m;][]{dem12} and, in the near future, by AzTEC on the LMT (1100~$\mu$m).  To put our technique in the context of these surveys, we simulate the counterpart identification results that can be expected with the improved positional accuracy of these observations.  We test two beamsizes: 14.5$^{\prime\prime}$, which is the approximate FWHM of SCUBA-2 at 850~$\mu$m, and 8.5$^{\prime\prime}$, the FWHM of AzTEC on the LMT in its early 32-m configuration.  The latter beamsize is also close to the FWHM of SCUBA-2 at 450~$\mu$m ($\sim$7$^{\prime\prime}$).  Additionally, we simulate 33$^{\prime\prime}$, the beamsize of AzTEC on ASTE and the COSMOS observations used in this study, which is comparable to the beamsize for SPIRE 500~$\mu$m observations.  

\begin{figure}
\includegraphics[scale=0.4, trim=15mm 5mm 10mm 0, clip]{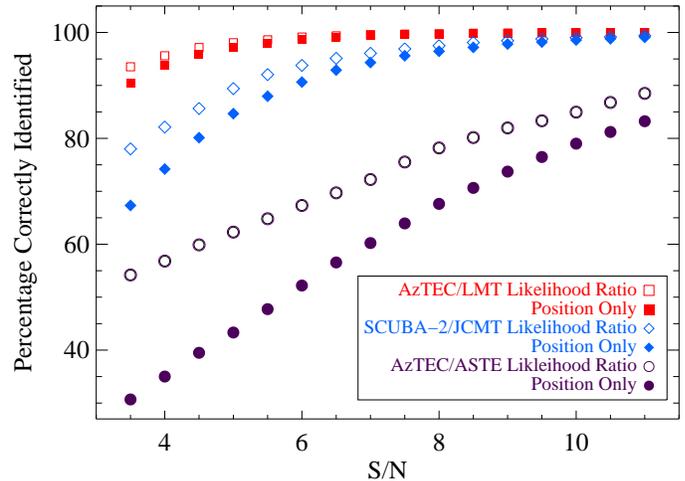}
\caption{The simulated percentage of correctly identified IRAC counterparts in COSMOS for beamsizes of 33$^{\prime\prime}$ (circles), 14.5$^{\prime\prime}$ (diamonds) and 8.5$^{\prime\prime}$ (squares), representing AzTEC on ASTE (this study), SCUBA-2 on JCMT, and AzTEC on LMT, respectively.  The results using this technique (empty symbols) are an improvement over using just positions (filled symbols) in the lowest S/N bins for all beamsizes.}
\label{fig:lmtsim}
\end{figure}

The simulations are done using the 68 SMGs in the COSMOS field that have a known radio and IRAC counterpart.  To simulate the effects of random noise in the mm observations, a random position is drawn for each SMG from a distribution given by Eqn~\ref{eqn:pu} which is centered on the radio counterpart and has a FWHM determined by the observations being simulated and a fixed S/N (Eqn~\ref{eqn:pu2}).  This random position determines the positional uncertainty, which is then combined with the IRAC colors of all candidate counterparts within an appropriate search radius as per the same blind-test performed in Section~\ref{sec:rank}.  This procedure is repeated 1,000 times for each SMG for different S/N bins.  The results of the simulation can be seen in Figure~\ref{fig:lmtsim}, contrasted with the results of using positional information only to predict the counterpart.  Our technique is an improvement over using only position in the lowest S/N bins in all cases, and as expected the amount of improvement is largest for increasing beamsize.  These simulations show that, even with higher resolution in future wide-area surveys, this technique can provide better predictions of SMG counterparts over positional coincidence, as well as a measure of robustness of the prediction through the reliability factor.  

%\subsection{Comparison with Previous Counterpart Identification Techniques}
%\label{sec:compare}

%As mentioned in the introduction, previous studies have mostly employed the P-statistic \citep{dow86} to find probable radio or MIPS counterparts.  In GOODS-N, \citet{cha09} reported 22/29 (76$\%$) robust counterparts using the 1.4 GHz radio or the MIPS 24$\mu$m P-statistic.  In GOODS-S, \citet{yun11} combines the radio and MIPS P-statistic with an IRAC color cut (S$_{\frac{3.6}{4.5}}$ $\ge$ 0.0) to identify 27/48 (56$\%$) robust counterparts based on at least on P-stat.  

\section{Summary and Conclusions}
\label{sec:con}

In this study, we have developed a new technique for multi-wavelength counterpart identification to submm detections by expanding on previous color cut techniques to create a characteristic density distribution of IRAC colors for the SMG population.  Our IRAC color-color distribution is based on a statistically significant set of 102 SMG counterparts determined through radio and SMA detections and built using kernel density estimation with cross-validation.  When combined with positional uncertainty information via the likelihood ratio, our technique provides a method for ranking IRAC galaxies surrounding SMGs according to their relative probability of being the correct counterpart.  In addition, we use the reliability factor to weigh the importance of a given likelihood ratio against the likelihood ratios of other candidate IRAC counterparts.

Once the IRAC color distribution is determined through known counterparts as in this study, it can then be applied to SMGs that lack other ancillary data, such as deep radio or MIPS 24~$\mu$m observations.  The strength of the CDD for SMGs stems from the ubiquitous nature of IRAC detections of SMGs and the discriminating power of the IRAC colors, which can be used as a redshift indicator for galaxy populations which 1) display a stellar bump in their infrared SEDs and 2) have a redshift distribution which peaks above $z$$\gtrsim$1.  Our overall technique, however, is general and can be applied to any population which occupies a unique color-magnitude or color-color space.  

In a blind-test using our training set, we show that our technique works well for mm observations with a moderate beamsize (e.g. 18$^{\prime\prime}$ in GOODS-N or 14.5$^{\prime\prime}$ for SCUBA-2 surveys) with an identification rate of 85 per cent, comparable to the current identification rate in GOODS-N, which utilizes extremely deep radio observations \citep{cha09}.  Our simulations demonstrate that this technique affords improvement over using only positional information, even for smaller beamsizes (e.g. 8.5$^{\prime\prime}$ for AzTEC on the LMT).  For larger beamsizes ($\gtrsim$30$^{\prime\prime}$), increased positional uncertainties and chance associations with high redshift non-SMG galaxies reduce the identification rate to 33-49 per cent.  These rates are comparable to current identification rates in COSMOS and GOODS-S \citep{are11,yun12}, using moderately deep radio observations.  Our technique, however, identifies a different sub-population of counterparts than previous radio studies by using IRAC data alone to determine counterparts.  These counterparts do not necessarily overlap with those identified through radio surveys and may have different properties, such as redshift distribution. Testing reveals no dependence of this method on the radio flux density, indicating it is applicable to radio-undetected SMGs.  Future expansions of this technique would include incorporating upper limits in order to allow for IRAC sources that are not detected in all four bands, and the expansion of the CDD beyond two dimensions to include additional information from a wider range of flux densities and colours.

Though the challenges put forth by submm counterpart identification will be eased in the near future by instruments and facilities such as SCUBA-2 on the JCMT, AzTEC on the LMT, and the relatively rapid interferometry of ALMA, there will still be a great need for practical and efficient counterpart identification techniques.   Single-dish facilities will increase the number of submm sources available for study by orders of magnitude within the next few years, making interferometric follow-up of each source impractical, while the need for multi-wavelength analysis remains vital for a complete picture of star forming galaxies and galaxy evolution.  While this study only addresses long-wavelength submm/mm observations, observations at shorter-wavelengths in the FIR/submm such as with BLAST or $Herschel$ suffer from similar counterpart identification challenges, indicating the continuing importance of counterpart identification studies.

\section*{Acknowledgements}

This work has been funded, in part, by NSF grant AST-0907952 and CONACyT projects CB-2009-133260 and CB-2011-167291.  This study is based, in part, on observations made with the $Spitzer$ Space Telescope, which is operated by the Jet Propulsion Laboratory at the California Institute of Technology under contract with NASA, and the National Radio Astronomy Observatory, which is a facility of the National Science Foundation operated under cooperative agreement by Associated Universities, Inc.  Kimberly S. Scott is supported by the National Radio Astronomy Observatory.  The James Clerk Maxwell Telescope is operated by the Joint Astronomy Centre on behalf of the Science and Technology Facilities Council of the United Kingdom, the Netherlands Organisation for Scientific Research, and the National Research Council of Canada.  The ASTE project is driven by Nobeyama Radio Observatory (NRO), a branch of National Astronomical Observatory of Japan (NAOJ), in collaboration with University of Chile, and Japanese institutes including University of Tokyo, Nagoya University, Osaka Prefecture University, Ibaraki University, Hokkaido University and Joetsu University of Education.

\label{bib}

\onecolumn

\appendix
\section{Data Tables}
\label{appendix:a}

%What data tables are needed for this paper? 

%\emph{AzTEC Source List}:  Is it necessary to report on the AzTEC source information?  This information is in other papers, but is only up-to-date in the COSMOS paper.  
%
%\emph{Radio-bright counterpart list (robust counterparts)}:  Could be like the GOODS-S paper with AzTEC ID, search radius, radio id/coordinates, distance, spitzer id/coodinates, likelihood ratio, reliability
%- include doubles and make robust counterparts in bold
%- list multiple IRAC galaxies?
%
%\emph{Radio-dim counterpart prediction list}:  List all possible IRAC galaxies and their reliabilities?  That would be hundreds.  
%
%\emph{Radio/IRAC/MIPS/other flux information list?}

\setlength\LTcapwidth{0.9\textwidth}
\small
\begin{longtable}{lcccccccc}
\caption{Training and Radio Counterpart Set:  this table contains the radio counterparts to AzTEC SMGs in GOODS-N, GOODS-S, and COSMOS.  The full table (385 rows) is available online.  Sources are included in the training set if the radio counterpart was detected in the original radio catalogs (non-extended, see Section~\ref{sec:multiples}) and if there is a corresponding IRAC detection to the radio counterpart that is detected in all four bands.   For COSMOS, radio detections which were found in the extended catalog (see Section~\ref{sec:multiples}) are indicated.    Three additional radio-undetected SMA-confirmed IRAC counterparts are listed and included in the training set.  For each source, the IRAC counterpart (if available) is listed on the same line as the radio counterpart, along with its rank, likelihood ratio, and reliability factor. The next two highest ranked IRAC galaxies are additionally listed (if available).  Some AzTEC sources in COSMOS are cross-listed with Table~\ref{app:multiples} due to having an additional nearby radio source in the extended radio catalog, making them multiple systems (see Section~\ref{sec:multiples}).  
 }\\
\hline
\multicolumn{9}{c}{} \\
AzTEC ID\tablenotemark{a} & $R_S$ & Radio Coords\tablenotemark{b} & P-stat$_{1.4\mbox{GHz}}$\tablenotemark{b} & IRAC Coords\tablenotemark{b} & Rank\tablenotemark{b} & Likelihood\tablenotemark{b} & Reliability\tablenotemark{b} & Other IDs\tablenotemark{c} \\
 & (") & (J2000) &  & (J2000) & & Ratio & Factor &  \\[10pt]
\hline
\endfirsthead
\hline
\multicolumn{9}{c}{} \\
AzTEC ID\tablenotemark{a} & $R_S$ & Radio Coords\tablenotemark{b} & P$_{1.4\mbox{GHz}}$\tablenotemark{b} & IRAC Coords\tablenotemark{b} & Rank\tablenotemark{b} & Likelihood & Reliability & Other IDs\tablenotemark{c} \\
 & (") & (J2000) &  & (J2000) & &  Ratio\tablenotemark{b} & Factor\tablenotemark{b} &  \\[10pt]
\hline
\multicolumn{9}{c}{} \\
\endhead
\multicolumn{9}{c}{} \\
\multicolumn{9}{c}{\underline{GOODS-N}} \\
\multicolumn{9}{c}{} \\
AzTEC/GN1 & 2.9 & {\bf J123711.89+622211.8} & {\bf 0.002} & {\bf J123711.88+622212.5} & {\bf 1/1} & {\bf 182.47}$\pm${\bf 4.7} & {\bf 1.00 } &  \\
 & & & &  $-$  &  $-$  &  $-$  &  $-$ &   \\
 & & & &  $-$  &  $-$  &  $-$  &  $-$ &  \\
AzTEC/GN3 & 5.2 & {\bf J123633.42+621408.7} & {\bf 0.001} & {\bf J123633.40+621408.8} & {\bf 1/4} & {\bf 67.17}$\pm${\bf 7.4} & 0.63 & GOODS850-5 \\
 & & & & J123633.25+621411.4 & 2/4 & 20.49$\pm$5.8 & 0.19 &  \\
\end{longtable}
\noindent{\scriptsize $^a$AzTEC IDs correspond to the IDs from \citet{dow11} for the GOODS fields and \citet{are11} for COSMOS.}  \\
{\scriptsize $^b$Radio counterparts with P-stat$\le$0.05 are bolded.  IRAC counterparts that are ranked correctly (1st) and } \\
{\scriptsize reliability factors greater than 0.8 are also bolded.} \\
{\scriptsize $^c$``AzTECxx" correspond to the IDs from \citet{you07} and \citet{you09} for SMA observations, ``COSLA-xx" correspond to IDs from \citet{smo12} for PdBI observations of AzTEC sources in the COSMOS field, and ``GOODS850-x" correspond to IDs from \citet{bar12} for SMA observations of SCUBA/AzTEC sources in GOODS-N. } \\
{\scriptsize $^d$Radio counterpart is identified in the extended 3$\sigma$ radio catalog (see Section~\ref{sec:multiples}).} \\
{\scriptsize $^e$AzTEC source is cross-listed with Table~\ref{app:multiples}.} \\
{\scriptsize $^f$IRAC counterpart is not detected in all four bands.} \\
{\scriptsize $^g$IRAC counterparts identified via the SMA \citep{you07, you09}.  These include two radio-undetected SMGs (AzTEC/C4 and AzTEC/C10) and one multiple system (AzTEC/C42), for which the SMA identified the correct counterpart and which is cross-listed with Table~\ref{app:multiples}.  Two SMA sources, AzTEC7 \citep{you07} and AzTEC10 \citep{you09} do not match any SMGs in our AzTEC catalogs and so their IRAC fluxes were taken from their respective papers. \\
{\scriptsize $^h$The true radio counterpart for AzTEC/GN14 is undetected in our radio catalogs.  The radio source listed here is unassociated with the submm emission \citep{bar12}.} 

\setlength\LTcapwidth{0.9\textwidth}
\small\begin{longtable}{lcccccccc}
\caption{Multiple Radio Set: List of SMGs in GOODS-N, GOODS-S, and COSMOS that have multiple radio detections.  The full table (108 rows) is available online.  The IRAC counterpart to each radio detection is listed on the same line with its rank, likelihood ratio, and reliability factor.  The information for an additional potential IRAC counterpart is often also listed.  For COSMOS, radio detections which were found in the extended catalog (see Section~\ref{sec:multiples}) are indicated.  Some AzTEC/COSMOS SMGs were single systems in the unextended radio catalog, but are multiple systems in the extended catalog.  These sources were used in the original training set and are cross-listed with Table~\ref{app:robust}.}\\
\hline
\multicolumn{9}{c}{} \\
AzTEC ID\tablenotemark{a} & $R_S$ & Radio Coords\tablenotemark{b} & P-stat$_{1.4\mbox{GHz}}$\tablenotemark{b} & IRAC Coords\tablenotemark{b} & Rank\tablenotemark{b} & Likelihood & Reliability & Other IDs\tablenotemark{c} \\
 & (") & (J2000) &  & (J2000) & & Ratio\tablenotemark{b} & Factor\tablenotemark{b} & \\[10pt]
\hline
\endfirsthead
\hline
\multicolumn{9}{c}{} \\
AzTEC ID\tablenotemark{a} & $R_S$ & Radio Coords\tablenotemark{b} & P$_{1.4\mbox{GHz}}$\tablenotemark{b} & IRAC Coords\tablenotemark{b} & Rank\tablenotemark{b} & Likelihood & Reliability \\
 & (") & (J2000) &  & (J2000) & &  Ratio\tablenotemark{b} & Factor\tablenotemark{b} \\[10pt]
\hline
\multicolumn{8}{c}{} \\
\endhead
\multicolumn{8}{c}{} \\
\multicolumn{8}{c}{\underline{GOODS-N}} \\
\multicolumn{8}{c}{} \\
AzTEC/GN7 & 6.4 & {\bf J123711.32+621330.9} & {\bf 0.028} & J123711.34+621331.4 & 3/4 & 21.40$\pm$0.4 & 0.17 \\
 & & {\bf J123711.99+621325.5} &  {\bf 0.030} & J123711.99+621326.0 & 1/4 & 48.23$\pm$2.4 & 0.39 \\
 & & & & J123711.86+621333.8 & 2/4 & 45.87$\pm$3.4 & 0.37 \\
AzTEC/GN11 & 7.4 & {\bf J123635.87+620707.8} & {\bf 0.008} & J123635.85+620708.0 & 1/8 & 10.09$\pm$0.1 & 0.37 \\
 & & {\bf J123635.87+620703.9} & {\bf 0.012} & $-$ & $-$ & $-$ & $-$ \\
 & & {\bf J123636.32+620706.8} & {\bf 0.034} & J123636.37+620707.3 & 3/8 & 5.31$\pm$0.1 & 0.20  \\
AzTEC/GN16 & 7.9 & {\bf J123615.94+621514.3} & {\bf 0.024} & J123615.82+621515.8 & 1/5 & 96.55$\pm$3.1 & 0.51 & GOODS850-7  \\
 & & {\bf J123616.10+621513.8} & {\bf 0.027} & J123616.10+621514.0\tablenotemark{d} & 2/5 & 74.93$\pm$1.2 & 0.40  \\
 & & & & J123616.84+621514.7 & 3/5 & 7.07$\pm$0.6 & 0.04   \\
AzTEC/GN17 & 8.0 & J123539.91+621430.8 & 0.071 & J123539.96+621431.1\tablenotemark{e} & $-$ & $-$ & $-$  \\
 & & J123539.93+621441.9 &  0.076 & J123539.94+621441.17\tablenotemark{e} & $-$ & $-$ & $-$  \\
 & & {\bf J123541.20+621438.5} &  {\bf 0.034} & J123541.25+621439.0\tablenotemark{e} &  $-$  & $-$  &  $-$  \\
AzTEC/GN18 & 8.0 & {\bf J123741.16+621220.5} &  {\bf 0.020} & J123741.16+621221.3 & 1/6 & 95.07$\pm$3.6 & 0.73   \\
 & & {\bf J123741.63+621223.6} & {\bf 0.047} & J123741.66+621224.0 & 2/6 & 21.91$\pm$1.2 & 0.17   \\
 & & & & J123740.12+621220.0 & 3/6 & 4.80$\pm$12.6 & 0.04   \\
AzTEC/GN24 & 8.7 & J123608.57+621435.3 &  0.058 & J123608.60+621435.7 & 1/6 & 33.28$\pm$1.5 & 0.50 \\
 & & J123608.64+621435.5 & 0.056 & J123608.60+621435.7\tablenotemark{d} & 1/6 & 33.28$\pm$1.5 & 0.50 \\
 & & & & J123607.66+621441.7 & 2/6 & 7.28$\pm$1.4 & 0.11   \\
AzTEC/GN30 & 9.1 & J123641.85+621719.7 & 0.068 & $-$ & $-$ & $-$ & $-$ \\
 & & {\bf J123642.18+621722.1} & {\bf 0.049} & J123642.19+621722.7 & 2/5 & 7.05$\pm$0.1 & 0.24   \\
 & & & & J123642.78+621712.5 & 1/5 & 8.98$\pm$4.8 & 0.30   \\
AzTEC/GN36 & 9.5 & {\bf J123617.39+621551.2} & {\bf 0.046} & J123617.44+621551.5 & 4/6 & 5.24$\pm$0.3 & 0.12  & GOODS850-3  \\
 & & {\bf J123617.54+621540.7} & {\bf 0.034} & J123617.55+621540.8 & 1/6 & 15.41$\pm$0.7 & 0.35   \\
 & & J123618.32+621550.5 &  0.094 & J123618.33+621550.8 & 3/6 & 7.62$\pm$0.3 & 0.17   \\
\multicolumn{9}{c}{} \\
\multicolumn{9}{c}{} \\
\multicolumn{9}{c}{\underline{GOODS-S}} \\
\multicolumn{9}{c}{} \\
AzTEC/GS19 & 10.3 & J033222.71-274127.0 & 0.080 & J033222.71-274126.2 & 3/7 & 4.26$\pm$0.2 & 0.13   \\
 & & J033222.92-274125.3 & 0.061 & J033222.85-274125.1 & 4/7 & 4.03$\pm$0.1 & 0.12   \\
 & & & & J033223.77-274131.6 & 1/7 & 11.66$\pm$0.7 & 0.34   \\
AzTEC/GS32 & 13.7 & J033310.14-275124.4 & 0.195 & J033310.12-275124.8 & 3/4 & 1.18$\pm$0.2 & 0.20   \\
 & & J033308.63-275134.5 & 0.092 & J033308.62-275134.5 & 1/4 & 2.03$\pm$0.1 & 0.34   \\
 & & & & J033308.78-275125.2 & 2/4 & 1.50$\pm$0.4 & 0.25   \\
AzTEC/GS35 & 12.6 & {\bf J033227.19-274051.5} & {\bf 0.004} & J033227.17-274051.7\tablenotemark{d} & 4/9 & 27.92$\pm$1.1 & 0.14   \\
 & & {\bf J033226.97-274049.9} &  {\bf 0.009} & J033227.17-274051.7 & 4/9 & 27.92$\pm$1.1 & 0.14   \\
 & & & & J033226.83-274056.1 & 1/9 & 50.66$\pm$8.3 & 0.26   \\
AzTEC/GS37 & 14.6 & J033255.70-274623.8 &  0.212 & $-$ & $-$ & $-$ & $-$ \\
 & & J033257.09-274624.5 &  0.150 & $-$ & $-$ & $-$ & $-$ \\
 & & & & {\bf J033256.71-274606.6} & {\bf 1/5} & {\bf 20.61$\pm$10.1} & {\bf 0.81}  \\
\multicolumn{9}{c}{} \\
\multicolumn{9}{c}{} \\
\multicolumn{9}{c}{\underline{COSMOS}} \\
\multicolumn{9}{c}{} \\
AzTEC/C23 & 11.0 & {\bf J100142.36+021836.0} & {\bf 0.013} & {\bf J100142.34+021835.2} & {\bf 1/2} & {\bf 25.74$\pm$1.9} & {\bf 0.90 } \\
 & & {\bf J100142.76+021841.5} & {\bf 0.036} & J100142.73+021841.1 & 2/2 & 2.62$\pm$0.2 & 0.09  \\
 & & & &  $-$  &  $-$  &  $-$ &  $-$  \\
AzTEC/C43\tablenotemark{f} & 12.5 & {\bf J100003.12+020201.5} & {\bf 0.032} & J100003.09+020201.5 & 5/8 & 7.91$\pm$2.3 & 0.05 \\
 & & J100003.76+020217.6\tablenotemark{g} & 0.069 & J100003.81+020216.7 & 7/8 & 1.25$\pm$0.1 & 0.01 \\
 & & & & J100003.40+020204.4 & 1/8 & 65.08$\pm$7.8 & 0.38 \\
AzTEC/C53 & 12.8 & {\bf J100122.47+021209.7}\tablenotemark{g} & {\bf 0.006} & J100122.46+021210.0 & 4/8 & 4.82$\pm$0.4 & 0.05 \\
 & & {\bf J100122.07+021212.5}\tablenotemark{g} & {\bf 0.038} & J100122.04+021213.9 & 2/8 & 1.39$\pm$0.3 & 0.02 \\
 & & & & J100122.61+021214.9 & 1/8 & 48.66$\pm$14.8 & 0.53 \\
AzTEC/C59 & 13.2 & {\bf J100030.13+023716.7} & {\bf 0.018} & J100030.14+023716.8 & 1/5 & 9.35$\pm$0.3 & 0.51 \\
 & & J100031.11+023717.2 & 0.066 & J100031.10+023717.2 & 4/5 & 1.44$\pm$0.0 & 0.08  \\
 & & & & J100303.32+014413.7 & 2/5 & 3.29$\pm$1.0 & 0.18  \\
AzTEC/C69 & 13.5 & {\bf J100138.00+020908.7}\tablenotemark{g} & {\bf 0.001} & J100138.00+020910.0 & 1/5 & 31.40$\pm$4.3 & 0.68 \\
 & & J100137.29+020904.6\tablenotemark{g} & 0.076 & J100137.35+020904.3 & 3/5 & 4.53$\pm$1.2 & 0.10   \\
 & & & & J100138.17+020913.3 & 2/5 & 8.27$\pm$1.6 & 0.18 \\
AzTEC/C70 & 13.5 & {\bf J100025.47+020312.6} & {\bf 0.037} & $-$ & $-$ & $-$ & $-$ \\
& & {\bf J100025.60+020316.1} & {\bf 0.022} & J100025.60+020316.2 & 3/4 & 3.19$\pm$0.1 & 0.09  \\
 & & & & J100025.94+020315.2 & 1/4 & 27.53$\pm$12.9 & 0.75  \\
AzTEC/C91\tablenotemark{f} & 14.5 & {\bf J100128.20+022340.8\tablenotemark{g}} & {\bf 0.038} & {\bf J100128.21+022340.8} & 3/8 & 2.22$\pm$0.1 & 0.07 \\
 & & {\bf J100128.48+022345.2} & {\bf 0.005} & J100128.48+022345.1 & 1/8 & 19.11$\pm$6.7 & 0.57  \\
 & & & & J100129.30+022348.2 & 2/8 & 7.33$\pm$0.5 & 0.22  \\
AzTEC/C100 & 14.7 & J095917.70+021028.8 & 0.091 & J095917.67+021028.5 & 4/7 & 1.14$\pm$0.1 & 0.02 \\
 & & J095917.98+021043.7 & 0.061 & J095917.96+021043.5 & 2/7 & 4.17$\pm$1.3 & 0.06  \\
 & & & & {\bf J095918.70+021040.9} & {\bf 1/7} & {\bf 61.71$\pm$10.5} & {\bf 0.86 }  \\
AzTEC/C110 & 14.9 & {\bf J100108.71+020036.4}\tablenotemark{g} & {\bf 0.001} & {\bf J100108.72+020036.6} & {\bf 1/5} & {\bf 35.27$\pm$2.6} & {\bf 0.88} \\
 & & {\bf J100108.50+020029.1}\tablenotemark{g} & {\bf 0.025} & J100108.46+020030.8 & 2/5 & 2.15$\pm$0.6 & 0.05   \\
 & & & & J100108.46+020030.8 & 3/5 & 2.15$\pm$0.6 & 0.05  \\
AzTEC/C124 & 15.3 & {\bf J095946.35+023602.1} & {\bf 0.033} & J095946.31+023602.3 & 3/6 & 1.91$\pm$0.1 & 0.09  \\
 & & {\bf J095946.68+023546.8} & {\bf 0.038} & J095946.63+023546.8 & 1/6 & 14.36$\pm$1.0 & 0.68  \\
 & & & & J095956.11+014253.5 & 2/6 & 2.43$\pm$0.9 & 0.12  \\
AzTEC/C140\tablenotemark{f} & 16.0 & {\bf J100124.96+015147.3} & {\bf 0.004} & {\bf J100125.00+015147.2} & 2/5 & 37.98$\pm$4.8 & 0.35  \\
 & & J100125.49+015135.38\tablenotemark{g} & 0.071 & J100125.47+015134.0\tablenotemark{d} & $-$ & $-$ & $-$ \\
 & & & & J100124.60+015141.3 & 1/5 & 61.43$\pm$9.8 & 0.57 \\
AzTEC/C142\tablenotemark{f} & 16.0 & J100018.82+020245.0 & 0.070 & J100018.83+020245.0 & 6/7 & 0.93$\pm$0.0 & 0.03  \\
 & & J100017.46+020252.6\tablenotemark{g} & 0.071 & J100017.47+020252.3 & 2/7 & 4.27$\pm$0.3 & 0.14   \\
 & & & & J100018.37+020245.1 & 1/7 & 18.97$\pm$14.7 & 0.61   \\
AzTEC/C149 & 16.2 & J100149.82+022831.9 & 0.051 & J100149.82+022831.9 & 4/9 & 2.13$\pm$0.2 & 0.06   \\
 & & {\bf J100150.92+022826.3} & {\bf 0.026} & J100150.92+022826.4 & 2/9 & 8.46$\pm$4.0 & 0.25   \\
 & & & & J100149.96+022824.5 & 1/9 & 16.85$\pm$9.2 & 0.49   \\
AzTEC/C150 & 16.2 & {\bf J100004.81+023045.1} & {\bf 0.016} & {\bf J100004.79+023045.3} & {\bf 1/5} & {\bf 80.30$\pm$1.6} & {\bf 0.91 }   \\
 & & J100005.43+023029.1 & 0.089 & J100005.44+023029.2 & 5/5 & 0.28$\pm$0.0 & 0.00   \\
 & & & & J100005.74+023039.1 & 2/5 & 3.68$\pm$1.0 & 0.04   \\
AzTEC/C154 & 16.3 & {\bf J100035.41+023531.8} & {\bf 0.046} & J100035.41+023532.1 & 4/10 & 4.39$\pm$0.4 & 0.09    \\
 & & {\bf J100035.84+023543.8} & {\bf 0.032} & J100035.85+023544.4 & 5/10 & 4.28$\pm$0.3 & 0.08   \\
 & &  & & J100035.39+023541.0 & 1/10 & 19.16$\pm$1.6 & 0.38  \\
AzTEC/C157\tablenotemark{f} & 16.4 & J100103.55+014810.7 & 0.059 & J100103.56+014810.8 & 2/9 & 27.02$\pm$0.8 & 0.33  \\
 & & {\bf J100104.79+014805.0}\tablenotemark{g} & {\bf 0.037} & J100104.76+014804.5 & 1/9 & 43.69$\pm$3.9 & 0.53 \\
 & & & & J100103.82+014815.1 & 3/9 & 5.49$\pm$4.0 & 0.07  \\
AzTEC/C158 & 16.5 & J100013.58+021230.8 & 0.104 & J100013.58+021230.8 & 8/8 & 0.06$\pm$0.0 & 0.00 \\
 & & J100015.28+021240.6 &  0.085 & J100015.28+021240.2 & 3/8 & 6.09$\pm$0.4 & 0.11  \\
 & & & & J100015.01+021235.2 & 1/8 & 29.52$\pm$24.0 & 0.54  \\
AzTEC/C169 & 16.9 & J100013.36+023016.2 &  0.125 & J100013.34+023016.0 & 7/8 & 0.28$\pm$0.0 & 0.01  \\
 & & J100014.21+023019.1 & 0.051 & {\bf J100014.21+023019.0} & {\bf 1/8} & {\bf 27.68$\pm$0.6} & {\bf 0.83}  \\
 & & & & J100013.79+023019.1 & 2/8 & 1.82$\pm$1.4 & 0.05  \\
AzTEC/C174\tablenotemark{f} & 17.0 & {\bf J100008.27+015623.9} &  {\bf 0.017} & J100008.28+015624.3 & 1/8 & 74.42$\pm$9.3 & 0.58  \\
 & & J100007.50+015629.7\tablenotemark{g} & 0.117 & J100007.54+015628.9 & 8/8 & 0.22$\pm$0.8 & 0.00  \\
 & & & & J100007.82+015614.4 & 2/8 & 32.73$\pm$5.1 & 0.25  \\
AzTEC/C178 & 17.1 & {\bf J100141.18+022002.2} & {\bf 0.010} & J100141.15+022002.0 & 1/10 & 15.68$\pm$1.0 & 0.46  \\
 & & J100142.30+022005.3 & 0.110 & J100142.29+022005.3 & 5/10 & 1.99$\pm$0.2 & 0.06  \\
 & & & & J100140.96+021957.7 & 2/10 & 9.98$\pm$13.3 & 0.29   \\
AzTEC/C181\tablenotemark{f} & 17.2 & {\bf J095929.55+021713.1} & {\bf 0.047} & $-$ & $-$ & $-$ & $-$ \\
 & & J095929.47+021658.1\tablenotemark{g} & 0.119 & J095929.47+021658.5 & $-$ & $-$ & $-$  \\
AzTEC/C183 & 17.3 & {\bf J100226.17+021227.1} & {\bf 0.00} & $-$ & $-$ & $-$ & $-$   \\
 & & {\bf J100226.25+021229.6} &  {\bf 0.003} & J100226.25+021228.8 & 1/11 & 7.92$\pm$0.5 & 0.24   \\
 & &  & & J100225.70+021224.2 & 2/11 & 6.81$\pm$3.5 & 0.20    \\
AzTEC/C184 & 17.3 & J100015.71+014445.6 &  0.054 & J100015.70+014445.5 & 5/10 & 1.89$\pm$0.1 & 0.04   \\
 & & J100016.33+014437.4 &  0.133 & J100016.34+014437.5 & 7/10 & 1.22$\pm$0.0 & 0.03   \\
 & &  {\bf J100016.77+014453.9}\tablenotemark{g}  & {\bf 0.044}  & J100016.81+014453.18 & 6/10 & 1.72$\pm$0.1 & 0.04   \\
 & & & & J100015.42+014447.3 & 1/10 & 15.83$\pm$4.3 & 0.37   \\
AzTEC/C187 & 17.4 & J100152.55+021954.3 & 0.079 & J100152.57+021954.0 & 2/13 & 14.97$\pm$0.3 & 0.28   \\
 & & J100153.34+021928.7 &  0.115 & J100153.34+021928.7 & 4/13 & 2.66$\pm$0.0 & 0.05  \\
 & & & & J100153.46+021939.6 & 1/13 & 20.67$\pm$5.0 & 0.39  \\
\multicolumn{9}{c}{} \\
\hline
\label{app:multiples}
\end{longtable}
\noindent{\scriptsize $^a$AzTEC IDs correspond to the IDs from \citet{dow11} for the GOODS fields and \citet{are11} for COSMOS.}  \\
{\scriptsize $^b$Radio detections with P-stat$\le$0.05 are bolded.  IRAC detections with reliability factors greater than 0.8 are also bolded.} \\
{\scriptsize $^c$``AzTECxx" correspond to the IDs from \citet{you07} and \citet{you09} for SMA observations, ``COSLA-xx" correspond to IDs from \citet{smo12} for PdBI observations of AzTEC sources in the COSMOS field, and ``GOODS850-x" correspond to IDs from \citet{bar12} for SMA observations of SCUBA/AzTEC sources in GOODS-N. } \\
{\scriptsize $^d$IRAC detection is within 2$^{\prime\prime}$ of both radio detections.} \\
{\scriptsize $^e$IRAC counterpart is not detected in all four bands.} \\
{\scriptsize $^f$AzTEC source is cross-listed with Table~\ref{app:robust}.} \\
{\scriptsize$^g$Radio counterpart is identified in the extended 3$\sigma$ radio catalog (see Section~\ref{sec:multiples}).} 

\setlength\LTcapwidth{0.9\textwidth}
\small
\begin{longtable}{lccccc}
\caption{Radio-Undetected SMG Set: list of SMGs without radio detections.  The full table (321 rows) is available online.  For each AzTEC source, the likelihood ratios and reliability factors for the top three ranked potential IRAC counterparts are listed.  IRAC detections with $R$ $\geq$ 0.8 are bolded.} \\
\hline
\multicolumn{6}{c}{} \\
AzTEC ID\tablenotemark{a} & $R_S$ &  IRAC Coords\tablenotemark{b} & Likelihood & Reliability & Other IDs \\
 & (") & (J2000) & Ratio\tablenotemark{b} & Factor\tablenotemark{b} &  \\[10pt]
\hline
\endfirsthead
\hline
\multicolumn{6}{c}{} \\
AzTEC ID\tablenotemark{a} & $R_S$ & IRAC Coords\tablenotemark{b} & Likelihood & Reliability & Other IDs\tablenotemark{c} \\
 & (") & (J2000) & Ratio\tablenotemark{b} & Factor\tablenotemark{b} &  \\[10pt]
\hline
\multicolumn{6}{c}{} \\
\endhead
\multicolumn{6}{c}{} \\
\multicolumn{6}{c}{\underline{GOODS-N}} \\
\multicolumn{6}{c}{} \\
AzTEC/GN2 & 4.8 & {\bf J123632.01+621711.1} & {\bf 15.26$\pm$2.1} & {\bf 0.99 }  \\
 & & $-$  &  $-$  &  $-$   \\
 & & $-$  &  $-$  &  $-$   \\
%AzTEC/GN7 & 6.4 & J123711.99+621326.0 & 48.23$\pm$2.4 & 0.39  \\
% & & J123711.86+621333.8 & 45.87$\pm$3.4 & 0.37  \\
% & & J123711.34+621331.4 & 21.40$\pm$0.4 & 0.17  \\
AzTEC/GN8 & 6.6 & J123645.85+621447.3 & 12.23$\pm$0.4 & 0.70 & GOODS850-11  \\
 & &  J123645.77+621442.6 & 5.05$\pm$0.6 & 0.29   \\
\end{longtable}
\noindent{\scriptsize $^a$AzTEC IDs correspond to the IDs from \citet{dow11} for the GOODS fields and \citet{are11} for COSMOS.}  \\
{\scriptsize $^b$IRAC detections with reliability factors greater than 0.8 are bolded.} \\
{\scriptsize $^c$``AzTECxx" correspond to the IDs from \citet{you07} and \citet{you09} for SMA observations, ``COSLA-xx" correspond to IDs from \citet{smo12} for PdBI observations of AzTEC sources in the COSMOS field, and ``GOODS850-x" correspond to IDs from \citet{bar12} for SMA observations of SCUBA/AzTEC sources in GOODS-N. } \\

\label{lastpage}

\end{document}